\documentclass[10pt,aps,prl,raggedbottom,longbibliography,nobalancelastpage,reprint,citeautoscript,letterpaper,superscriptaddress]{revtex4-2}
\usepackage{graphicx}
\usepackage{mathrsfs}
\usepackage{amsfonts}
\usepackage{times}
\usepackage{amsmath}
\usepackage{color}
\usepackage[colorlinks,linkcolor=blue,citecolor=blue]{hyperref}
\usepackage{mathtools}
\usepackage{orcidlink}
\usepackage{multirow}
\usepackage{bm}

\newcommand{\T}{$\mathcal{T}$}
\newcommand{\PT}{$\mathcal{PT}$}

\usepackage{ulem}

\begin{document}
\title{Multiferroic Collinear Antiferromagnet with Hidden Altermagnetic Split}

\author{Jin Matsuda}
\email{jin-matsuda@g.ecc.u-tokyo.ac.jp}
\affiliation{Department of Applied Physics, The University of Tokyo, Hongo, Bunkyo-ku, Tokyo 113-8656, Japan}

\author{Hikaru Watanabe}
\email{hikaru-watanabe@g.ecc.u-tokyo.ac.jp}
\affiliation{Department of Physics, The University of Tokyo, Hongo, Bunkyo-ku, Tokyo 113-0033, Japan}

\author{Ryotaro Arita}
\affiliation{Department of Physics, The University of Tokyo, Hongo, Bunkyo-ku, Tokyo 113-0033, Japan}
\affiliation{RIKEN Center for Emergent Matter Science, 2-1 Hirosawa, Wako, Saitama 351-0198, Japan}
\begin{abstract}
Altermagnets exhibit nonrelativistic spin splitting due to the breaking of time-reversal symmetry and have been garnering significant attention as promising materials for spintronic applications.
In contrast, conventional antiferromagnets without spin splitting seem not to have any symmetry breaking and have drawn less attention.
However, we show that conventional antiferromagnets with a nonzero propagation vector ($\bm{Q}$ vector) bring about nontrivial symmetry breakings.
The incompatibility between the $\bm{Q}$ vector and nonsymmorphic symmetry leads to macroscopic symmetry breaking without lifting spin degeneracy. 
Moreover, the hidden altermagnetic spin splitting in the electronic structure gives rise to various emergent responses.
To examine our prediction, we perform first-principles calculations for MnS$_2$ and investigate its multiferroic properties, such as nonlinear transport and optical activity. 
Our findings reveal unique properties in conventional antiferromagnets, providing another perspective for designing spintronic materials.
\end{abstract}

\maketitle

\noindent \textit{Introduction---}
Spintronics is a rapidly advancing field that seeks to harness the spin degree of freedom in electronic devices. 
A key focus in this area has been spin-orbit coupling (SOC), an essential ingredient of spin-related responses. 
Recently, a new class of collinear magnets, dubbed altermagnets~\cite{Smejkal2022-zq, Smejkal2022-uk}, has garnered significant attention. 
Altermagnets are collinear antiferromagnets breaking the time-reversal ($\mathcal{T}$) symmetry and exhibit nonrelativistic spin splitting (NRSS) with $d$-wave or higher-order even-parity symmetries in the electronic structure~\cite{Lopez-Moreno2012-ov, Noda2016-vj, Naka2019-az, Hayami2019-le, Ahn2019-lt, Yuan2020-qf, Smejkal2020-wl, Yuan2021-hq, Mazin2021-vi, Egorov2021-ey, Smejkal2022-zq, Smejkal2022-uk}. 
This unique type of spin-charge coupling originates from exchange splitting energy and can reach magnitudes as high as electron volt in certain materials~\cite{Smejkal2022-uk}.
Since the scale of this exchange energy is much larger than that of SOC, altermagnets are significantly broadening the range of materials available for spintronic applications.
Recent studies have identified various spin-related phenomena that altermagnets can host~\cite{Smejkal2020-wl, Ahn2019-lt, Naka2019-az, Gonzalez-Hernandez2021-kj, Zhou2024-oa, Smejkal2022-dz,Tanaka2024-eu,Tanaka2024-il, Fernandes2024-db, Bhowal2024-ty} and have proposed numerous candidate materials~\cite{Reimers2024-tx, Krempasky2024-db, Takagi2024-om, Reichlova2024-fo, Osumi2024-lb, Lee2024-jk}. 

Altermagnet allows for only the even-parity spin splitting due to the symmetry unique to its collinear spin arrangement~\footnote{
Let us write the spin-momentum-locking texture of the electronic bands in collinear antiferromagnets as $F(\bm{k})S_z$ with the form factor $F (\bm{k})$ and $z$-component of spin $S_z$.
Applying the $[-1 \cdot C_2 ||1]$ symmetry ($-1$ in the spin space means the \T{} operation) to $F(\bm{k})S_z$ yields $F(\bm{k}) = F(-\bm{k})$~\cite{Smejkal2022-uk}.
}.
Spin splitting can be more diverse in noncollinear magnetic materials. 
For example, the odd-parity spin-splitting texture arises in $p$-wave magnets driven by noncollinear spin arrangements~\cite{Sivianes2024-hk, Brekke2024-zk, Hellenes2023-ro}. 
The parity-odd spin-momentum coupling results in phenomena not found in altermagnets~\cite{Pari2024-mv, Chakraborty2024-xy}.
On the other hand, spin order can realize emergent responses even without NRSS.
For instance, the parity-violating but $\mathcal{PT}$-symmetric magnets~\cite{Watanabe2024-qo}, such as Cr$_2$O$_3$ and CuMnAs, are known to exhibit effects that couple electric fields with spin, \textit{e.g.}, electromagnetic responses~\cite{Dzyaloshinskii1959-gm, Astrov1960-kw, Rado1961-tg} , electrical switching~\cite{Wadley2016-rx,Bodnar2018-jg} and electrically-tunable NRSS~\cite{Yuan2023-wv, Guo2024-pl}, despite of the $\mathcal{PT}$-enforced spin degeneracy.

Moreover, spin space group, an extension of magnetic space group to the nonrelativistic limit~\cite{Litvin1974-qv, Litvin1977-ti, Brinkman1966-si, Jiang2024-oz, Chen2024-td, Xiao2024-im, Watanabe2024-mk, Shinohara2024-pe}, has revealed that various responses can emerge due to unique symmetries, even in spin-degenerate materials~\cite{Watanabe2024-mk}.
This behavior can be attributed to spin-translation symmetry, symmetry for the combination of spin-space rotation and lattice translation.
NRSS can be prohibited by spin translation symmetry even without $\mathcal{PT}$ symmetry~\cite{Martin2008-ze, Feng2020-ag}. However, spin-translation symmetry does not restrict the occurrence of the anomalous Hall effect~\cite{Takagi2023-kk} and nonlinear transport~\cite{Watanabe2024-mk,Zhu2024-re}. As a result, these responses can be finite even in the absence of NRSS.
In the spin space group framework, symmetry operations are written as $[g_\text{sp}||g_\text{orb}]$, where $g_\text{sp}$ acts in the spin space and $g_\text{orb}$ acts in the real space. 
For example, $[C_2||\bm{t}]$ represents a twofold rotation of spins ($C_2$)  and lattice translation ($\bm{t}$). 
In collinear spin structures, $g_\text{sp} = C_2$ flips spins as in the case of the $\mathcal{T}$ operation but keeps the electron's momentum $\bm{k}$, and thus the spin translation symmetry $[C_2||\bm{t}]$ protects the spin degeneracy of the electronic structure.
The symmetry for $[C_2||\bm{t}]$ is present only in spin configurations with the nonzero propagation vector $\bm{Q} \neq \bm{0}$ at a time-reversal-invariant momentum.
We assume such $\bm{Q}$ vectors in the following.

\begin{figure}
    \begin{center}
        \includegraphics[width=8.5cm, clip]{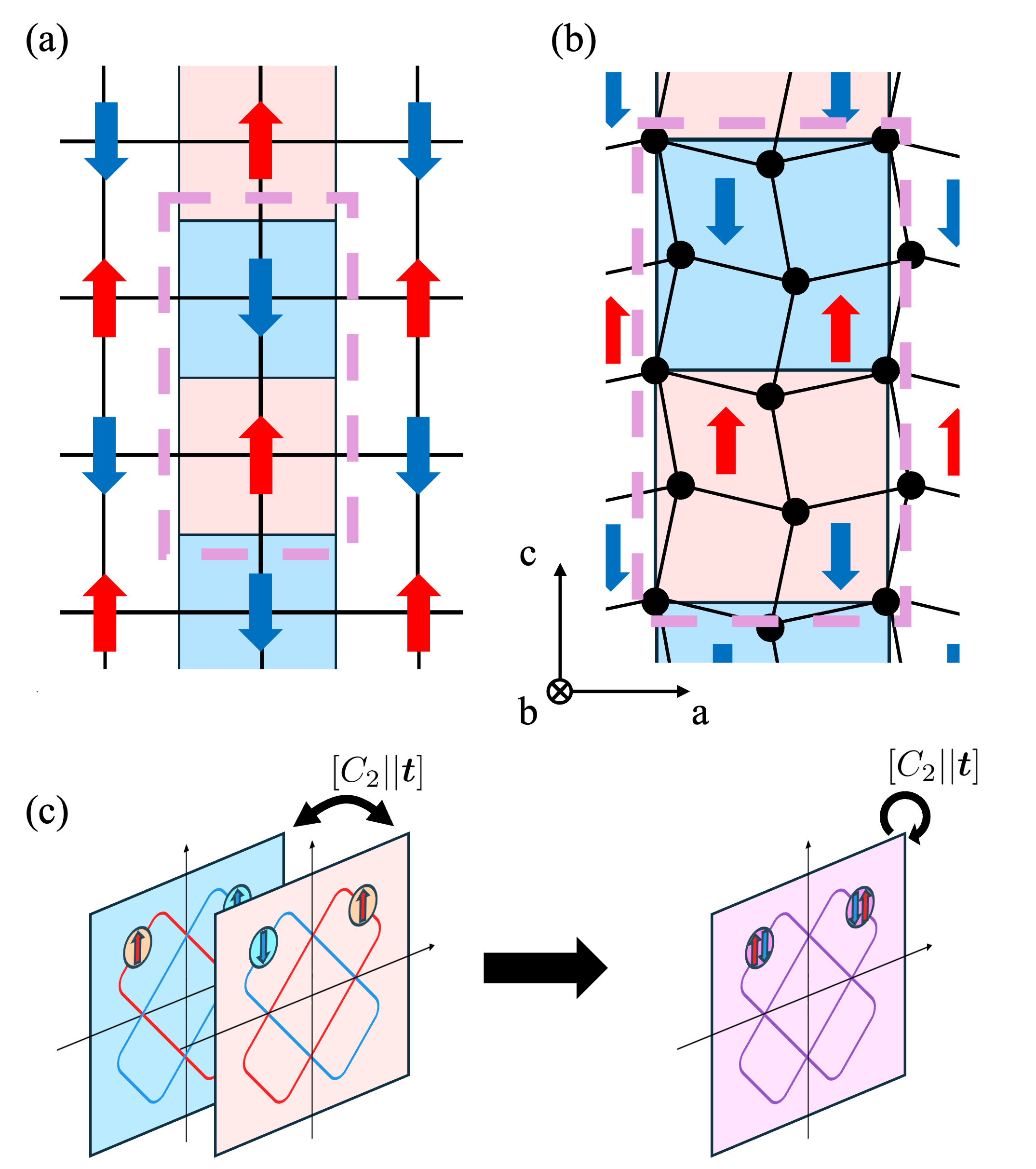}
    \end{center}
    \caption{
 The role of a finite $\bm{Q}$ vector in collinear antiferromagnets.
  (a) The checkerboard-patterned antiferromagnetic order with $\bm{Q} \neq 0$.
   The red and blue squares represent the original unit cell, and the purple dashed rectangle represents the magnetic unit cell.
 (b) The Rutile-like crystal structure showing the $bc$-glide symmetry and its collinear spin order with a finite $\bm{Q}$ vector.
 The incompatibility between a nonsymmorphic symmetry and a finite $\bm{Q}$ vector results in the violation of the parity symmetry whose inversion centers are the magnetic sites.
 (c) Hidden spin polarization in the electronic structure of collinear antiferromagnet depicted in panel (b).
 (left panel) In each of the original unit cells (blue and red squares), nonrelativistic $d$-wave spin splitting is allowed due to the absence of the $\mathcal{PT}$ and $[C_2||\bm{t}]$ symmetries within each sector.
 (right panel) The $[C_2||\bm{t}]$ symmetry in the magnetic unit cell (the purple square) leads to spin degeneracy at each momentum~\cite{Yuan2023-wv}.
 }
    \label{fig:hidden_spin}
\end{figure}

The spin order with the nonzero $\bm{Q}$ vector breaks the microscopic translation symmetry, inflating the unit cell without violating the spin degeneracy at each momentum. Nonetheless, it may significantly affect the electronic property.
For instance, the $[C_2||\bm{t}]$ symmetry implies that electrons localized at each of the two original unit cells may undergo the staggered NRSS.
It points to the hidden spin polarization driven by the spin order [FIG.~\ref{fig:hidden_spin}], which can be sizable since it stems from the Coulomb interaction.
Leveraging this property could pave the way for innovative functionalities of antiferromagnetic materials.

In this Letter, we propose a new class of magnets with hidden spin polarization, that is, multiferroic antiferromagnet whose inversion ($\mathcal{P}$) symmetry is broken by collinear spin ordering in a SOC-free manner.
A candidate magnet has no NRSS due to its spin translation symmetry and is thereby classified as a conventional antiferromagnet~\cite{Smejkal2022-uk}.
The nonsymmorphic property coupled to the $\bm{Q}$ vector, however, nontrivially breaks the macroscopic symmetry, leading to the SOC-free physical consequences such as nonlinear transport and optical activity.
This work highlights intriguing responses in conventional antiferromagnets and offers another perspective for designing functional magnets.

\noindent \textit{Nonsymmorphic symmetry with finite $\bm{Q}$ vector---}
Hidden spin polarization can be found in diverse collinear antiferromagnets with a finite $\bm{Q}$ vector.
A prototypical example is the checkerboard-patterned antiferromagnetic order in a square lattice, which can be found in cuprates [FIG.~\ref{fig:hidden_spin}\,(a)].
The staggered spin polarization induces the $s$-wave splitting in each sublattice, but it does not realize emergent responses due to no change in the point group symmetry.
It is, however, not the case in the nonsymmorphic crystals. 

We illustrate the nontrivial symmetry violation by taking the two-dimensional net mimicking the rutile crystal structure [FIG.~\ref{fig:hidden_spin}(b)].
The paramagnetic state shows a glide symmetry in the $bc$ plane, of which the half translation $\bm{t}/2$ along $\bm{c}$ connects neighboring unit cells.
For the spin order with the $\bm{Q}$ vector depicted in FIG.~\ref{fig:hidden_spin}(b), the collinear spin arrangements in red- and blue-shaded original unit cells are respectively characterized by the spin Laue group hosting the $d$-wave spin splitting [FIG.~\ref{fig:hidden_spin}(c)]
~\footnote{
    The spin Laue group describes the point group symmetry of SOC-free collinear magnets enhanced by the real-space inversion symmetry~\cite{Smejkal2022-zq}.
    After dividing the overall spin point group by the spin-only group~\cite{Litvin1974-qv}, the nontrivial spin point group is obtained as $\mathcal{P} = \bm{H} \cup [C_2 || A] \bm{H}$ where the group $\bm{H}$ consists of only the real-space operations.
    The spin Laue group is obtained as $\bm{H} = mmm$ and $A = C_{4b}$ if one assumes the zero-$\bm{Q}$ spin order in FIG.~\ref{fig:hidden_spin}\,(b).
    The obtained group allows the $d$-wave spin splitting, implying the altermagnetic spin splitting within red- and blue-shaded cells in FIG.~\ref{fig:hidden_spin}\,(b).
    Note that the $bc$ glide operation allowing for a mirror symmetry of $\bm{H}$ is not kept due to the finite-$\bm{Q}$ spin order coupled to the nonsymmorphic property.
}.
The overall altermagnetic split is perfectly compensated by the time-reversal operation combined with the translation $\bm{t}$ as well as the spin translation symmetry $[C_2 || \bm{t}]$~\cite{Yuan2023-wv}.
Then, the spin degeneracy in the electronic structure is kept at every momentum.

The $\bm{Q}$ vector parallel to the half translation, however, makes the glide operation incompatible with the spin order inflating the unit cell.
Similarly to the glide operation, the space-inversion symmetry is broken, indicating the (magnetoelectric) multiferroicity without NRSS~\cite{Perez-Mato2016-qa}.
The multiferroic state undergoes sizable changes in its electronic structure by the hidden altermagnetic split.

The resultant macroscopic symmetry breaking can be identified by the orbital point group defined by
        \begin{equation}
            P_\text{orb}= \{ \text{det}\, g_\text{sp} \cdot R_\text{orb} \big| ~[g_\text{sp} || (R_\text{orb}, \bm{t})] \in \mathcal{G} \},
            \label{orbital-point-group}
        \end{equation}
where $R_\text{orb}$ denotes the point-group operation acting on the real space, and $\mathcal{G}$ is the spin space group of a given magnetic material.
An improper spin-space rotation leads to $\text{det}\, g_\text{sp} =-1$ denoting the time-reversal operation in the real space, which is denoted by $1'$ in the magnetic-point-group notation.
The orbital point group is useful for identifying the spin-order-induced emergent phenomena regarding the orbital degree of freedom.
We note that the orbital point group is a supergroup of the magnetic point group where SOC is taken into account.
In the case of FIG.~\ref{fig:hidden_spin}(b), the orbital-point-group symmetry is reduced from $P_\text{orb} = 4/mmm1'$ of the paramagnetic state to $P_\text{orb} = 2mm1'$ with $C_{2a}$ of the ordered state.

\begin{figure}
    \begin{center}
        \includegraphics[width=8.5cm, clip]{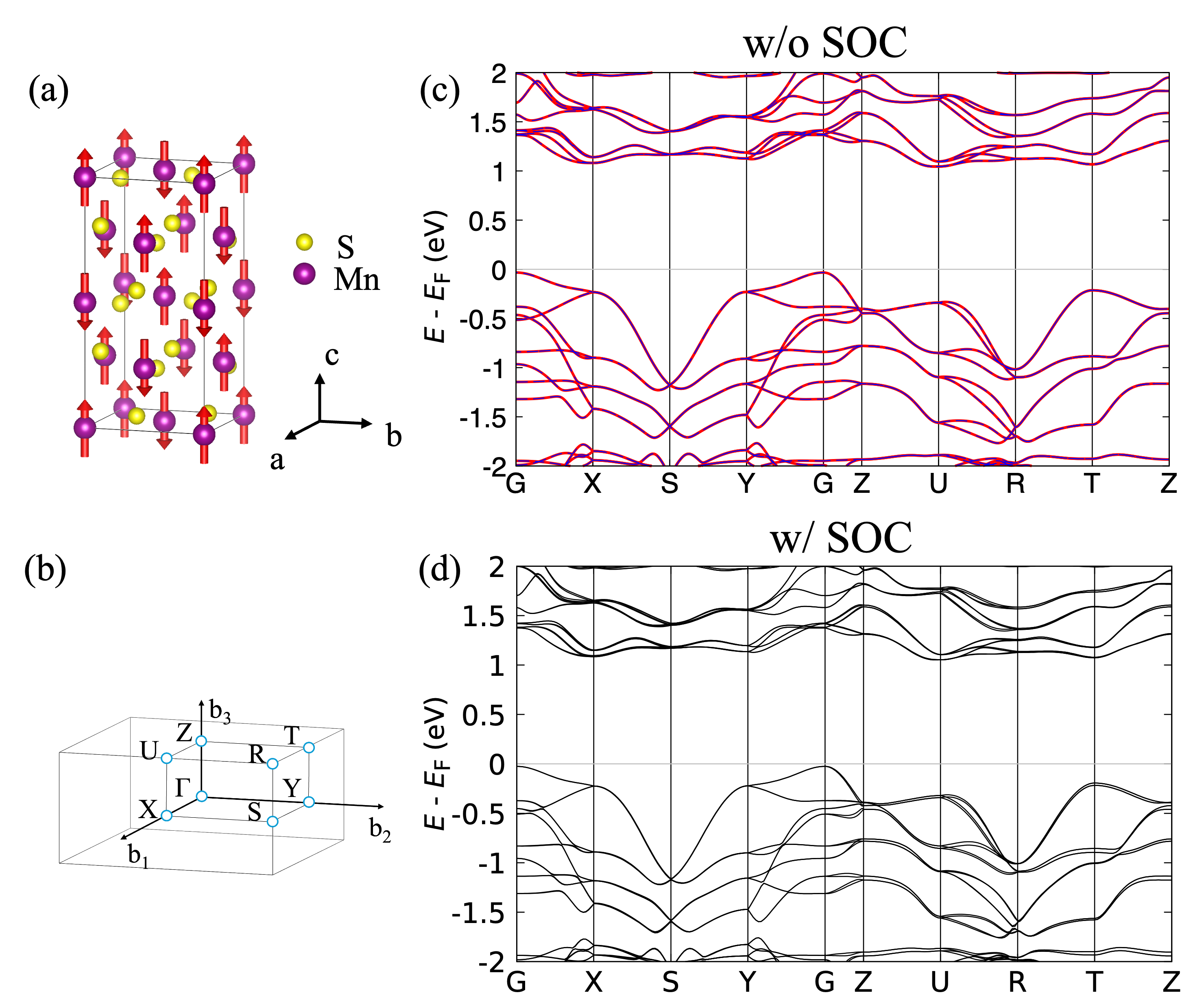}
    \end{center}
    \caption{
 (a) Crystal and magnetic structures of MnS$_2$.
 (b) Brillouin zone.
 (c) The band structure of MnS$_2$ without SOC.
 The red solid line represents the up spin band, and the blue dotted line represents the down spin band.
 (d) The band structure of MnS$_2$ with SOC.}
    \label{fig:MnS2}
\end{figure}

\noindent \textit{Nonlinear transport and optical activity of} MnS$_2${---}
Let us demonstrate our proposal by first-principles calculations of MnS$_2$, a collinear antiferromagnet with a finite $\bm{Q}$ vector.
Figure~\ref{fig:MnS2}\,(a) depicts the crystal and magnetic structures of MnS$_2$~\cite{Corliss1958-yi}, which exhibit collinear antiferromagnetic order and retains the $\mathcal{T}$ symmetry.
In its paramagnetic phase, the space group of MnS$_2$ is $Pa\bar{3}$, and MnS$_2$ possesses a glide plane in the $ac$-axis.
However, the spin order breaks the $ac$-glide and $\mathcal{P}$ symmetries.
As a result, the orbital point group $P_\text{orb}$ is reduced from a cubic centrosymmetric group $m\bar{3}1'$ to a polar group $m2m1'$.

The calculations are carried out with the Vienna \textit{Ab-Initio} Simulation Package (VASP)~\cite{Kresse1996-oj} with the Perdew-Burke-Ernzerhof (PBE) exchange-correlation functional~\cite{Blochl1994-bi, Perdew1996-hz}.
The crystal and magnetic structures are obtained from those reported in Ref.~\cite{Corliss1958-yi}.
To reproduce the experimental band gap of 1 eV~\cite{Brostigen1970-vh}, we make use of the DFT + U method, with the Hubbard parameter $U = 3$ eV and $J = 1$ eV for Mn-3$d$ orbitals~\cite{Anisimov1991-na,Persson2006-tx}.
For nonrelativistic calculations, the plane-wave energy cutoff is 400 eV, and the $\Gamma$-centered $8 \times 8 \times 8$ $k$-point mesh is employed for self-consistent calculations.
The Wannier tight-binding Hamiltonian constructed with Wannier90 code~\cite{Souza2001-at, Pizzi2020-gd} consists of 88 Wannier functions to incorporate Mn-3$d$ and S-2$p$ orbitals.
\textsc{postw.x} module is adopted for calculations of the Berry curvature, Berry curvature dipole (BCD), and interband gyration tensor $\gamma_{abc}$~\cite{Wang2006-sn, Tsirkin2018-ly, Malashevich2010-vt}.
The result with SOC is obtained with the plane-wave energy cutoff 500 eV and $8 \times 8 \times 4$ $k$ points.
The tight-binding Hamiltonian with the SOC effect is obtained with 176 Wannier functions, including Mn-3$d$ and S-2$p$ orbitals.
$100^3$ $k$ points are taken for evaluating BCD and gyration tensor.

Figure~\ref{fig:MnS2}\,(c) is for the band structure calculated without SOC.
Notably, the energy bands remain spin-degenerate, consistent with the characteristics of conventional antiferromagnets with the spin translation symmetry $[C_2||\bm{t}]$.
Slight Rashba spin splitting appears by the SOC effect since $\mathcal{PT}$ and $[C_2||\bm{t}]$ are absent in the case with SOC [FIG.~\ref{fig:MnS2}\,(d)].
The SOC modification to the electronic bands, however, is not significant, being consistent with the weak SOC of MnS$_2$.

\begin{figure*}
    \begin{center}
        \includegraphics[width=17.5cm, clip]{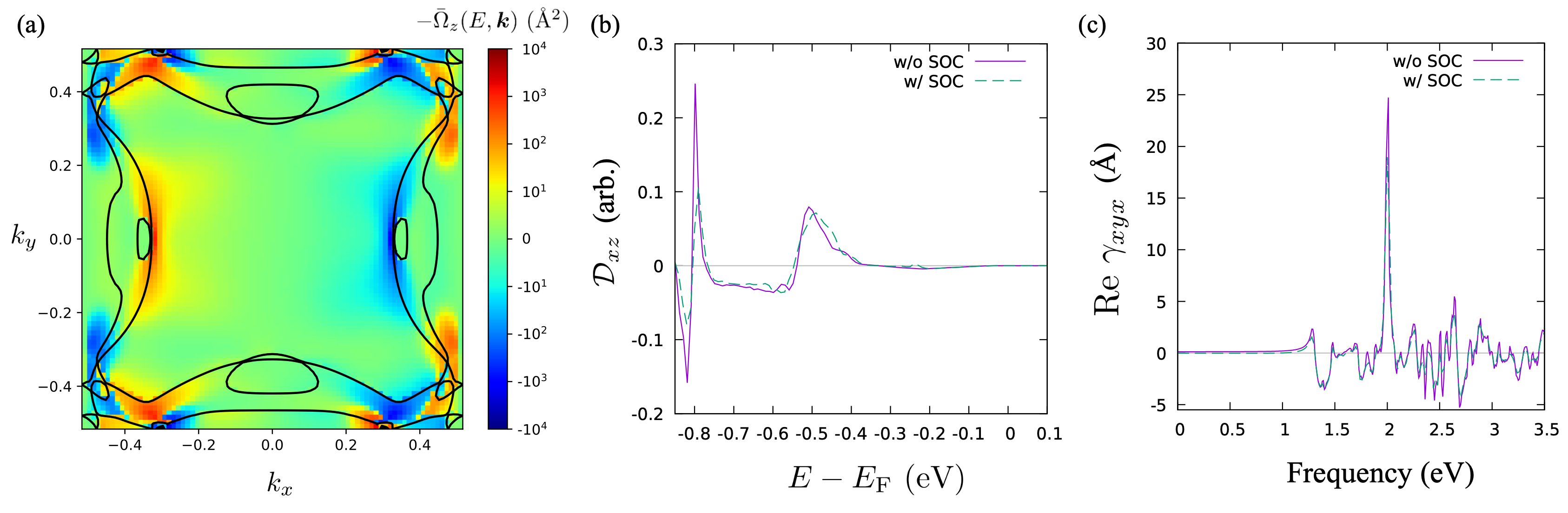}
    \end{center}
    \caption{Properties of MnS$_2$ stemming from the parity breaking.
 (a) Distribution of the $z$ component of the Berry curvature summed up to $E-E_{\mathrm{F}} = -2$~eV in the plane $k_z = 0$.
 (b) Reference-energy ($E$) dependence of Berry curvature dipole $\mathcal{D}_{xz}(E)$.
 The solid line represents the value without SOC, and the dashed line represents that with SOC.
 (c) The interband gyration tensor and its dependence on the frequency of the irradiating light.
 The solid line is for the case without SOC, while the dashed line is for that with SOC.}
    \label{fig:MnS2_properties}
\end{figure*}

We demonstrate the multiferroic property of the MnS$_2$ by the Berry curvature arising from the spin ordering. 
Figure~\ref{fig:MnS2_properties}\,(a) illustrates the $z$-component of the Berry curvature $\bar{\Omega}_z (E,\bm{k})$ in the $k_z=0$ plane.
$\bar{\Omega}_z (E,\bm{k})$ is the summation of the $\bm{k}$-local Berry curvature of $n$-th band over the bands below the energy $E$ as $\bar{\Omega}_z (E,\bm{k}) = \sum_{n}^{E_n <E} \Omega_{z n} (\bm{k})$.
The $\mathcal{PT}$-symmetry breaking by the finite-$\bm{Q}$ spin ordering allows for a nonzero Berry curvature even without NRSS.
The momentum distribution of $\Omega_z (\bm{k})$ clearly shows the $p$-wave anisotropy as adapted to its $\mathcal{T}$-symmetric and polar symmetry.
A typical value of the Berry curvature of MnS$_2$ is comparable to that significantly observed in the body-centered-cubic Fe~\cite{Wang2006-sn}.

The Berry curvature can be sizable in total as quantified by BCD.
In FIG.~\ref{fig:MnS2_properties}\,(b),  we plot the BCD defined by
\begin{equation}
    \mathcal{D}_{ab}(E) = \int_{\mathrm{BZ}} \frac{d^3 \bm{k}}{(2\pi)^3} \sum_n \frac{\partial E_n}{\partial k_a} \Omega_n^b(\bm{k}) \left(-\frac{\partial f_0}{\partial \varepsilon}\right)_{\varepsilon = E_n}, 
\end{equation}
where $a, b \in \{x, y, z\}$, and $f_0 = \Theta(\varepsilon - E)$, with a step function $\Theta (\cdot)$.
$\mathcal{D}_{xz}$ and $\mathcal{D}_{zx}$ are nonzero in MnS$_2$.
The BCD is unique to the $\mathcal{T}$-symmetric and noncentrosymmetric systems and contributes to nonlinear phenomena such as the Hall effect and photocurrent generation~\cite{Deyo2009-ah,Moore2010-sy,Sodemann2015-vl}.
Consistent with the SOC-free nature of multiferroic symmetry, the BCD attains large values even in the absence of SOC and NRSS and does not undergo a notable modification of SOC.

Moreover, the multiferroic property of MnS$_2$ can be detected by the natural rotatory power $\rho_0^{\mathrm{inter}} (\omega)$ given by~\cite{Malashevich2010-vt, Tsirkin2018-ly}
\begin{equation}
    \rho_0^{\mathrm{inter}} (\omega) = \frac{\omega^2}{2c_0^2} \mathrm{Re}~\gamma_{abc}^{\mathrm{inter}} (\omega),
\end{equation}
which is determined by the interband gyration tensor $\gamma_{abc}^\text{inter}$ in insulators ($\omega$ is frequency of light, $c_0$ is the speed of light).
The definition of $\gamma_{abc}^{\mathrm{inter}}$ is in Supplemental Material~\cite{supple}.
The interband gyration tensor of MnS$_2$ is shown in FIG.~\ref{fig:MnS2_properties}\,(c).
Only $\gamma_{xyx}$ and $\gamma_{yzz}$ are nonzero by the symmetry.
Remarkably, the interband gyration tensor emerges in MnS$_2$ without NRSS and does not differ with and without SOC, similarly to BCD.
Taking $\mathrm{Re}~\gamma_{xyx} \simeq 0.12$ at $\omega = 0.12$ eV of the CO$_2$ laser light, we estimate the natural rotatory power to be $\rho_0 \simeq 0.023 $ (rad/cm).
Also, in the case of $\omega = 1$ eV, the natural rotatory power is significant as large as $\rho_0 \simeq 2.55$ (rad/cm).
For selenium, the natural rotatory power is approximately $5.24$ (rad/cm) at $\omega \simeq 1$ eV~\cite{Newnham2004-sd}, indicating that the spin-order-induced parity breaking can be comparable to that originated from the noncentrosymmetric crystal structure of selenium.

\noindent \textit{Discussions---}
This study delved into collinear antiferromagnets from a viewpoint different from that of altermagnets.
We presented the classification of collinear antiferromagnets in terms of the $\mathcal{T}$ and $\mathcal{PT}$ symmetries as summarized in TABLE~\ref{symm-class}.
In the classification concerning NRSS~\cite{Smejkal2022-zq}, the \PT{}-symmetric or \T{}-symmetric antiferromagnet is labeled conventional due to no NRSS, but they can be further classified by their parity under the \T{} operation.
Notably, the \T{}-symmetric antiferromagnets are realized by the spin order with the finite $\bm{Q}$ vector, while the altermagnets and \PT{}-symmetric magnets exist with either of the zero or nonzero $\bm{Q}$ vector~\cite{Jaeschke-Ubiergo2024-jc}.
Thus, \T{}-symmetric collinear antiferromagnets can be called \textit{Q magnets}.

\begin{table}
    \caption{Classification of collinear antiferromagnets.
             $\mathcal{PT}$-symmetric magnets and Q-magnets belong to conventional antiferromagnets in the classification by nonrelativistic spin splitting~\cite{Smejkal2022-uk}.}
    \begin{ruledtabular}
    \begin{tabular}{ccccc}
            & $\bm{Q}$ vector & NRSS &  $\mathcal{T}$ & $\mathcal{PT}$ \\\colrule
        Altermagnets &  $\bm{Q} = 0 $ / $\bm{Q} \neq 0$& $\checkmark$ & $\times$ & $\times$  \\
        $\mathcal{PT}$-symmetric magnets & $\bm{Q} = 0$ / $\bm{Q} \neq 0$& $\times$ & $\times$ & $\checkmark$ \\
        Q-magnets & $\bm{Q} \neq 0$ & $\times$ & $\checkmark$ & $\checkmark$/$\times$
    \end{tabular}
    \label{symm-class}
    \end{ruledtabular}
\end{table}

Q magnets cover a broad range of collinear magnets including that showing the nontrivial symmetry violation as we demonstrated with MnS$_2$, though they are less versatile in terms of the bulk spin transport due to no NRSS.
Further characterization of Q magnets can be done by the use of the orbital point group.
One can find a trivial example such as MnBi$_2$Te$_4$~\cite{Ding2020-gf}, which crystallizes in a symmorphic space group $R\bar{3}m$.
The magnetic phase of MnBi$_2$Te$_4$ is characterized by the orbital point group $P_\text{orb} = \bar{3}m1'$, which is identical to that of the paramagnetic phase and hence does not lead to emergent phenomena related to the orbital degree of freedom.
As a result, the nonsymmorphic property is pivotal for functional Q magnets manifesting nontrivial orbital-symmetry violations.
More details of the characterization are in Supplemental Material~\cite{supple}.

The orbital point group further implies the classification of magnetic materials without SOC, covering collinear and noncollinear magnets, in terms of the odd-parity NRSS and (spin-order-induced) $\mathcal{P}$ breaking of $P_\text{orb}$.
We identify three classes: (I) magnets without odd-parity NRSS but with the orbital $\mathcal{P}$ violation, (II) with odd-parity NRSS but without the orbital $\mathcal{P}$ violation, and (III) with both of two features.
Note that three classes are merged by SOC since the odd-parity NRSS and orbital $\mathcal{P}$ violation are always coupled to each other through it.
Class I covers materials of our main interests like MnS$_2$.
Class III consists of noncollinear magnets such as TbMn$_2$O$_5$~\cite{Blake2005-of}.
Spin order giving rise to the polar symmetry in $P_\text{orb}$ can be found in Class I and III, and the obtained spin order is switchable by the electric field.

\begin{figure}
    \begin{center}
        \includegraphics[width=0.90\linewidth, clip]{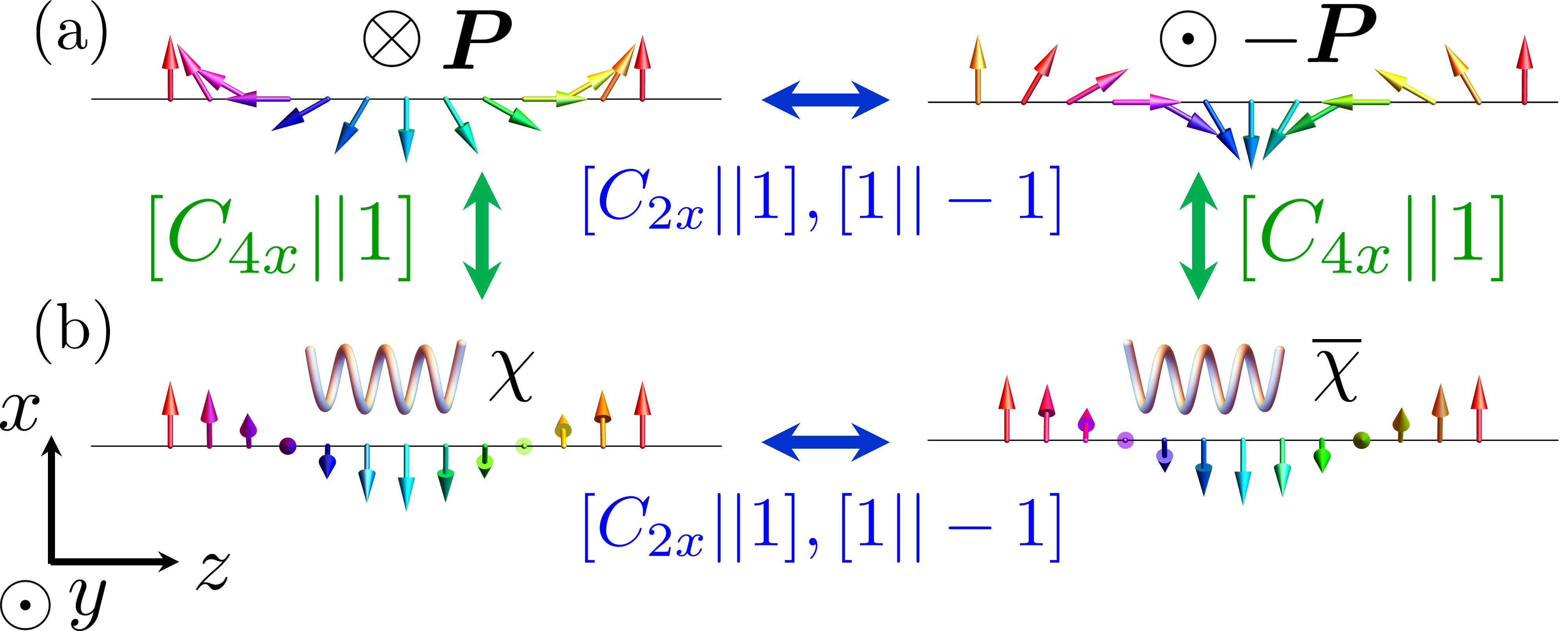}
    \end{center}
    \caption{
        Incommensurate spin textures and their relation by a symmetry operation $[g_\text{sp} || g_\text{orb}]$.
        (a) Spiral spin textures with the opposite swirling, which induce electric polarization $\pm \bm{P} \parallel \hat{b}$ through SOC.
        (b) Helical spin textures with the opposite swirling coupled to the crystal chirality $\chi$, $\overline{\chi}$ through SOC.
        The oppositely-swirling textures for each of the spiral and helical cases are related by the spin-space two-fold rotation ($g_\text{sp} = C_{2x}$) or the orbital-space parity operation ($g_\text{orb} = -1$), shown by the blue two-headed arrows.
        Similarly, the helical and spiral textures are related by the spin-space four-fold rotation ($g_\text{sp} = C_{4x}$), depicted by the green two-headed arrows.
    }
    \label{fig:incommensurate}
\end{figure}

To corroborate Class II, let us consider incommensurate magnets such as spiral and helical magnets where the Rashba- and chiral-type spin-momentum couplings show up in the SOC-free manner, respectively.
When the SOC effect is taken into account, the spiral spin order is coupled to the electric polarization $\bm{P}$ perpendicular to the $\bm{Q}$ vector~\cite{Kimura2003-ma,Katsura2005-fj}, while the helical order is coupled to the chirality $\chi$~\cite{Jiang2020-dr} [FIG.~\ref{fig:incommensurate}].
On the other hand, the oppositely-swirling spin textures are transformed into each other by $[C_{2x}|| 1]$ or $[1 || -1]$, revealing the $[C_{2x} || -1]$ symmetry for each configuration.
This intact symmetry points to the $\mathcal{P}$ symmetry in their orbital point groups forbiddening $\bm{P}$ and $\chi$, because the proper spin rotation ($\text{det}~C_{2x} = 1$) is neglected as in Eq.~\eqref{orbital-point-group}.
Thus, the magnets of Class II can show activity related to the spin degree of freedom by its odd-parity NRSS but are inactive with respect to the orbital degree of freedom.
This is also the case for commensurate magnets; \textit{e.g.}, an anti-perovskite Ce$_3$NIn shows noncoplanar spin order giving rise to the $p$-wave NRSS~\cite{Gabler2008-uy,Hellenes2023-ro}, whereas its orbital point group is centrosymmetric $P_\text{orb} = 4/mmm1'$.
Instead, one can manipulate the magnetic domains by making use of the odd-parity spin-momentum coupling; \textit{e.g.}, magnetic state as well as $p$-wave NRSS can be switched by applying the electric current $\bm{J}$ and magnetic field $\bm{B}$ as recently demonstrated by helimagnets~\cite{Jiang2020-dr,Jiang2021-yl,Ohe2021-cb,Masuda2024-tz}.
The switchable property can be understood by the correspondence between the $p$-wave NRSS $k_a S_b$ and the product of fields $J_a B_b$.
We note that SOC activates the orbital $\mathcal{P}$-symmetry breaking.

Moreover, two types of odd-parity NRSS arising from the incommensurate spin ordering are equivalent due to the arbitrariness of the spin-space frame.
One can confirm that the spiral and helical spin textures are related by $[C_{4x}|| 1]$.
Then, we conclude that the shape of odd-parity NRSS does not necessarily specify the type of the orbital $\mathcal{P}$ violation as one does with accounting for SOC~\cite{Watanabe2018-do,Hayami2018-bh}.
As a result, the proposed classification of the odd-parity magnets deepens our understanding of the role of SOC in the spin multiferroics~\cite{Tokura2014-ix}.

To summarize, we have investigated Q magnets, $\mathcal{T}$ symmetric collinear antiferromagnets with a finite $\bm{Q}$ vector. 
Q magnets are classified as conventional antiferromagnets~\cite{Smejkal2022-uk}, since they lack both NRSS and net magnetization.
However, the interplay between nonsymmorphic symmetry and a finite $\bm{Q}$ vector induces multiferroic responses without SOC or NRSS.
Furthermore, their responses can be switched by an electric field due to their polar symmetry originating from a finite $\bm{Q}$ vector.
Our proposal further broadens the search for responses in collinear antiferromagnets, offering a new perspective different from the framework of altermagnetism.

\textit{Acknowledgments---}
The authors thank Satoru Hayami, Rikuto Oiwa, and Mitsuaki Kawamura for their invaluable comments.
This work is supported by Grant-in-Aid for Scientific Research from JSPS, KAKENHI Grant No.~JP23K13058 (H.W.), No.~JP24K00581 (H.W.), No.~JP21H04990 (R.A.), JST-CREST No.~JPMJCR23O4 (R.A.), JST-ASPIRE No.~JPMJAP2317 (R.A.).
This work was supported by the RIKEN TRIP initiative (RIKEN Quantum, Advanced General Intelligence for Science Program, Many-body Electron Systems).
The crystal and magnetic structures are visualized by \textsc{VESTA}~\cite{Momma2011-jl}.


\begin{thebibliography}{84}%
    \makeatletter
    \providecommand \@ifxundefined [1]{%
     \@ifx{#1\undefined}
    }%
    \providecommand \@ifnum [1]{%
     \ifnum #1\expandafter \@firstoftwo
     \else \expandafter \@secondoftwo
     \fi
    }%
    \providecommand \@ifx [1]{%
     \ifx #1\expandafter \@firstoftwo
     \else \expandafter \@secondoftwo
     \fi
    }%
    \providecommand \natexlab [1]{#1}%
    \providecommand \enquote  [1]{``#1''}%
    \providecommand \bibnamefont  [1]{#1}%
    \providecommand \bibfnamefont [1]{#1}%
    \providecommand \citenamefont [1]{#1}%
    \providecommand \href@noop [0]{\@secondoftwo}%
    \providecommand \href [0]{\begingroup \@sanitize@url \@href}%
    \providecommand \@href[1]{\@@startlink{#1}\@@href}%
    \providecommand \@@href[1]{\endgroup#1\@@endlink}%
    \providecommand \@sanitize@url [0]{\catcode `\\12\catcode `\$12\catcode `\&12\catcode `\#12\catcode `\^12\catcode `\_12\catcode `\%12\relax}%
    \providecommand \@@startlink[1]{}%
    \providecommand \@@endlink[0]{}%
    \providecommand \url  [0]{\begingroup\@sanitize@url \@url }%
    \providecommand \@url [1]{\endgroup\@href {#1}{\urlprefix }}%
    \providecommand \urlprefix  [0]{URL }%
    \providecommand \Eprint [0]{\href }%
    \providecommand \doibase [0]{https://doi.org/}%
    \providecommand \selectlanguage [0]{\@gobble}%
    \providecommand \bibinfo  [0]{\@secondoftwo}%
    \providecommand \bibfield  [0]{\@secondoftwo}%
    \providecommand \translation [1]{[#1]}%
    \providecommand \BibitemOpen [0]{}%
    \providecommand \bibitemStop [0]{}%
    \providecommand \bibitemNoStop [0]{.\EOS\space}%
    \providecommand \EOS [0]{\spacefactor3000\relax}%
    \providecommand \BibitemShut  [1]{\csname bibitem#1\endcsname}%
    \let\auto@bib@innerbib\@empty
    \bibitem [{\citenamefont {Šmejkal}\ \emph {et~al.}(2022{\natexlab{a}})\citenamefont {Šmejkal}, \citenamefont {Sinova},\ and\ \citenamefont {Jungwirth}}]{Smejkal2022-zq}%
      \BibitemOpen
      \bibfield  {author} {\bibinfo {author} {\bibfnamefont {L.}~\bibnamefont {Šmejkal}}, \bibinfo {author} {\bibfnamefont {J.}~\bibnamefont {Sinova}},\ and\ \bibinfo {author} {\bibfnamefont {T.}~\bibnamefont {Jungwirth}},\ }\bibfield  {title} {\bibinfo {title} {Beyond conventional ferromagnetism and antiferromagnetism: A phase with nonrelativistic spin and crystal rotation symmetry},\ }\href {https://link.aps.org/doi/10.1103/PhysRevX.12.031042} {\bibfield  {journal} {\bibinfo  {journal} {Phys. Rev. X}\ }\textbf {\bibinfo {volume} {12}},\ \bibinfo {pages} {031042} (\bibinfo {year} {2022}{\natexlab{a}})}\BibitemShut {NoStop}%
    \bibitem [{\citenamefont {Šmejkal}\ \emph {et~al.}(2022{\natexlab{b}})\citenamefont {Šmejkal}, \citenamefont {Sinova},\ and\ \citenamefont {Jungwirth}}]{Smejkal2022-uk}%
      \BibitemOpen
      \bibfield  {author} {\bibinfo {author} {\bibfnamefont {L.}~\bibnamefont {Šmejkal}}, \bibinfo {author} {\bibfnamefont {J.}~\bibnamefont {Sinova}},\ and\ \bibinfo {author} {\bibfnamefont {T.}~\bibnamefont {Jungwirth}},\ }\bibfield  {title} {\bibinfo {title} {Emerging research landscape of altermagnetism},\ }\href {https://link.aps.org/doi/10.1103/PhysRevX.12.040501} {\bibfield  {journal} {\bibinfo  {journal} {Phys. Rev. X}\ }\textbf {\bibinfo {volume} {12}},\ \bibinfo {pages} {040501} (\bibinfo {year} {2022}{\natexlab{b}})}\BibitemShut {NoStop}%
    \bibitem [{\citenamefont {López-Moreno}\ \emph {et~al.}(2012)\citenamefont {López-Moreno}, \citenamefont {Romero}, \citenamefont {Mejía-López}, \citenamefont {Muñoz},\ and\ \citenamefont {Roshchin}}]{Lopez-Moreno2012-ov}%
      \BibitemOpen
      \bibfield  {author} {\bibinfo {author} {\bibfnamefont {S.}~\bibnamefont {López-Moreno}}, \bibinfo {author} {\bibfnamefont {A.~H.}\ \bibnamefont {Romero}}, \bibinfo {author} {\bibfnamefont {J.}~\bibnamefont {Mejía-López}}, \bibinfo {author} {\bibfnamefont {A.}~\bibnamefont {Muñoz}},\ and\ \bibinfo {author} {\bibfnamefont {I.~V.}\ \bibnamefont {Roshchin}},\ }\bibfield  {title} {\bibinfo {title} {First-principles study of electronic, vibrational, elastic, and magnetic properties of {FeF}$_2$ as a function of pressure},\ }\href {http://link.aps.org/pdf/10.1103/PhysRevB.85.134110} {\bibfield  {journal} {\bibinfo  {journal} {Phys. Rev. B}\ }\textbf {\bibinfo {volume} {85}},\ \bibinfo {pages} {134110} (\bibinfo {year} {2012})}\BibitemShut {NoStop}%
    \bibitem [{\citenamefont {Noda}\ \emph {et~al.}(2016)\citenamefont {Noda}, \citenamefont {Ohno},\ and\ \citenamefont {Nakamura}}]{Noda2016-vj}%
      \BibitemOpen
      \bibfield  {author} {\bibinfo {author} {\bibfnamefont {Y.}~\bibnamefont {Noda}}, \bibinfo {author} {\bibfnamefont {K.}~\bibnamefont {Ohno}},\ and\ \bibinfo {author} {\bibfnamefont {S.}~\bibnamefont {Nakamura}},\ }\bibfield  {title} {\bibinfo {title} {Momentum-dependent band spin splitting in semiconducting {MnO}$_2$: a density functional calculation},\ }\href {http://dx.doi.org/10.1039/c5cp07806g} {\bibfield  {journal} {\bibinfo  {journal} {Phys. Chem. Chem. Phys.}\ }\textbf {\bibinfo {volume} {18}},\ \bibinfo {pages} {13294} (\bibinfo {year} {2016})}\BibitemShut {NoStop}%
    \bibitem [{\citenamefont {Naka}\ \emph {et~al.}(2019)\citenamefont {Naka}, \citenamefont {Hayami}, \citenamefont {Kusunose}, \citenamefont {Yanagi}, \citenamefont {Motome},\ and\ \citenamefont {Seo}}]{Naka2019-az}%
      \BibitemOpen
      \bibfield  {author} {\bibinfo {author} {\bibfnamefont {M.}~\bibnamefont {Naka}}, \bibinfo {author} {\bibfnamefont {S.}~\bibnamefont {Hayami}}, \bibinfo {author} {\bibfnamefont {H.}~\bibnamefont {Kusunose}}, \bibinfo {author} {\bibfnamefont {Y.}~\bibnamefont {Yanagi}}, \bibinfo {author} {\bibfnamefont {Y.}~\bibnamefont {Motome}},\ and\ \bibinfo {author} {\bibfnamefont {H.}~\bibnamefont {Seo}},\ }\bibfield  {title} {\bibinfo {title} {Spin current generation in organic antiferromagnets},\ }\href {http://dx.doi.org/10.1038/s41467-019-12229-y} {\bibfield  {journal} {\bibinfo  {journal} {Nat. Commun.}\ }\textbf {\bibinfo {volume} {10}},\ \bibinfo {pages} {4305} (\bibinfo {year} {2019})}\BibitemShut {NoStop}%
    \bibitem [{\citenamefont {Hayami}\ \emph {et~al.}(2019)\citenamefont {Hayami}, \citenamefont {Yanagi},\ and\ \citenamefont {Kusunose}}]{Hayami2019-le}%
      \BibitemOpen
      \bibfield  {author} {\bibinfo {author} {\bibfnamefont {S.}~\bibnamefont {Hayami}}, \bibinfo {author} {\bibfnamefont {Y.}~\bibnamefont {Yanagi}},\ and\ \bibinfo {author} {\bibfnamefont {H.}~\bibnamefont {Kusunose}},\ }\bibfield  {title} {\bibinfo {title} {Momentum-dependent spin splitting by collinear antiferromagnetic ordering},\ }\href {https://doi.org/10.7566/JPSJ.88.123702} {\bibfield  {journal} {\bibinfo  {journal} {J. Phys. Soc. Jpn.}\ }\textbf {\bibinfo {volume} {88}},\ \bibinfo {pages} {123702} (\bibinfo {year} {2019})}\BibitemShut {NoStop}%
    \bibitem [{\citenamefont {Ahn}\ \emph {et~al.}(2019)\citenamefont {Ahn}, \citenamefont {Hariki}, \citenamefont {Lee},\ and\ \citenamefont {Kuneš}}]{Ahn2019-lt}%
      \BibitemOpen
      \bibfield  {author} {\bibinfo {author} {\bibfnamefont {K.-H.}\ \bibnamefont {Ahn}}, \bibinfo {author} {\bibfnamefont {A.}~\bibnamefont {Hariki}}, \bibinfo {author} {\bibfnamefont {K.-W.}\ \bibnamefont {Lee}},\ and\ \bibinfo {author} {\bibfnamefont {J.}~\bibnamefont {Kuneš}},\ }\bibfield  {title} {\bibinfo {title} {Antiferromagnetism in {RuO$_2$} as $d$-wave {Pomeranchuk} instability},\ }\href {http://link.aps.org/pdf/10.1103/PhysRevB.99.184432} {\bibfield  {journal} {\bibinfo  {journal} {Phys. Rev. B}\ }\textbf {\bibinfo {volume} {99}},\ \bibinfo {pages} {184432} (\bibinfo {year} {2019})}\BibitemShut {NoStop}%
    \bibitem [{\citenamefont {Yuan}\ \emph {et~al.}(2020)\citenamefont {Yuan}, \citenamefont {Wang}, \citenamefont {Luo}, \citenamefont {Rashba},\ and\ \citenamefont {Zunger}}]{Yuan2020-qf}%
      \BibitemOpen
      \bibfield  {author} {\bibinfo {author} {\bibfnamefont {L.-D.}\ \bibnamefont {Yuan}}, \bibinfo {author} {\bibfnamefont {Z.}~\bibnamefont {Wang}}, \bibinfo {author} {\bibfnamefont {J.-W.}\ \bibnamefont {Luo}}, \bibinfo {author} {\bibfnamefont {E.~I.}\ \bibnamefont {Rashba}},\ and\ \bibinfo {author} {\bibfnamefont {A.}~\bibnamefont {Zunger}},\ }\bibfield  {title} {\bibinfo {title} {Giant momentum-dependent spin splitting in centrosymmetric low- {Z} antiferromagnets},\ }\href {http://link.aps.org/pdf/10.1103/PhysRevB.102.014422} {\bibfield  {journal} {\bibinfo  {journal} {Phys. Rev. B}\ }\textbf {\bibinfo {volume} {102}},\ \bibinfo {pages} {014422} (\bibinfo {year} {2020})}\BibitemShut {NoStop}%
    \bibitem [{\citenamefont {Šmejkal}\ \emph {et~al.}(2020)\citenamefont {Šmejkal}, \citenamefont {González-Hernández}, \citenamefont {Jungwirth},\ and\ \citenamefont {Sinova}}]{Smejkal2020-wl}%
      \BibitemOpen
      \bibfield  {author} {\bibinfo {author} {\bibfnamefont {L.}~\bibnamefont {Šmejkal}}, \bibinfo {author} {\bibfnamefont {R.}~\bibnamefont {González-Hernández}}, \bibinfo {author} {\bibfnamefont {T.}~\bibnamefont {Jungwirth}},\ and\ \bibinfo {author} {\bibfnamefont {J.}~\bibnamefont {Sinova}},\ }\bibfield  {title} {\bibinfo {title} {Crystal time-reversal symmetry breaking and spontaneous {Hall} effect in collinear antiferromagnets},\ }\href {https://www.science.org/doi/10.1126/sciadv.aaz8809} {\bibfield  {journal} {\bibinfo  {journal} {Sci. Adv.}\ }\textbf {\bibinfo {volume} {6}},\ \bibinfo {pages} {eaaz8809} (\bibinfo {year} {2020})}\BibitemShut {NoStop}%
    \bibitem [{\citenamefont {Yuan}\ \emph {et~al.}(2021)\citenamefont {Yuan}, \citenamefont {Wang}, \citenamefont {Luo},\ and\ \citenamefont {Zunger}}]{Yuan2021-hq}%
      \BibitemOpen
      \bibfield  {author} {\bibinfo {author} {\bibfnamefont {L.-D.}\ \bibnamefont {Yuan}}, \bibinfo {author} {\bibfnamefont {Z.}~\bibnamefont {Wang}}, \bibinfo {author} {\bibfnamefont {J.-W.}\ \bibnamefont {Luo}},\ and\ \bibinfo {author} {\bibfnamefont {A.}~\bibnamefont {Zunger}},\ }\bibfield  {title} {\bibinfo {title} {Prediction of low-{Z} collinear and noncollinear antiferromagnetic compounds having momentum-dependent spin splitting even without spin-orbit coupling},\ }\href {http://link.aps.org/pdf/10.1103/PhysRevMaterials.5.014409} {\bibfield  {journal} {\bibinfo  {journal} {Phys. Rev. Mater.}\ }\textbf {\bibinfo {volume} {5}},\ \bibinfo {pages} {014409} (\bibinfo {year} {2021})}\BibitemShut {NoStop}%
    \bibitem [{\citenamefont {Mazin}\ \emph {et~al.}(2021)\citenamefont {Mazin}, \citenamefont {Koepernik}, \citenamefont {Johannes}, \citenamefont {González-Hernández},\ and\ \citenamefont {Šmejkal}}]{Mazin2021-vi}%
      \BibitemOpen
      \bibfield  {author} {\bibinfo {author} {\bibfnamefont {I.~I.}\ \bibnamefont {Mazin}}, \bibinfo {author} {\bibfnamefont {K.}~\bibnamefont {Koepernik}}, \bibinfo {author} {\bibfnamefont {M.~D.}\ \bibnamefont {Johannes}}, \bibinfo {author} {\bibfnamefont {R.}~\bibnamefont {González-Hernández}},\ and\ \bibinfo {author} {\bibfnamefont {L.}~\bibnamefont {Šmejkal}},\ }\bibfield  {title} {\bibinfo {title} {Prediction of unconventional magnetism in doped {FeSb$_2$}},\ }\href {https://www.pnas.org/doi/abs/10.1073/pnas.2108924118} {\bibfield  {journal} {\bibinfo  {journal} {Proc. Natl. Acad. Sci. U. S. A.}\ }\textbf {\bibinfo {volume} {118}},\ \bibinfo {pages} {e2108924118} (\bibinfo {year} {2021})}\BibitemShut {NoStop}%
    \bibitem [{\citenamefont {Egorov}\ and\ \citenamefont {Evarestov}(2021)}]{Egorov2021-ey}%
      \BibitemOpen
      \bibfield  {author} {\bibinfo {author} {\bibfnamefont {S.~A.}\ \bibnamefont {Egorov}}\ and\ \bibinfo {author} {\bibfnamefont {R.~A.}\ \bibnamefont {Evarestov}},\ }\bibfield  {title} {\bibinfo {title} {Colossal spin splitting in the monolayer of the collinear antiferromagnet {MnF$_2$}},\ }\href {https://pubs.acs.org/doi/abs/10.1021/acs.jpclett.1c00282} {\bibfield  {journal} {\bibinfo  {journal} {J. Phys. Chem. Lett.}\ }\textbf {\bibinfo {volume} {12}},\ \bibinfo {pages} {2363} (\bibinfo {year} {2021})}\BibitemShut {NoStop}%
    \bibitem [{\citenamefont {González-Hernández}\ \emph {et~al.}(2021)\citenamefont {González-Hernández}, \citenamefont {Šmejkal}, \citenamefont {Výborný}, \citenamefont {Yahagi}, \citenamefont {Sinova}, \citenamefont {Jungwirth},\ and\ \citenamefont {Železný}}]{Gonzalez-Hernandez2021-kj}%
      \BibitemOpen
      \bibfield  {author} {\bibinfo {author} {\bibfnamefont {R.}~\bibnamefont {González-Hernández}}, \bibinfo {author} {\bibfnamefont {L.}~\bibnamefont {Šmejkal}}, \bibinfo {author} {\bibfnamefont {K.}~\bibnamefont {Výborný}}, \bibinfo {author} {\bibfnamefont {Y.}~\bibnamefont {Yahagi}}, \bibinfo {author} {\bibfnamefont {J.}~\bibnamefont {Sinova}}, \bibinfo {author} {\bibfnamefont {T.}~\bibnamefont {Jungwirth}},\ and\ \bibinfo {author} {\bibfnamefont {J.}~\bibnamefont {Železný}},\ }\bibfield  {title} {\bibinfo {title} {Efficient electrical spin splitter based on nonrelativistic collinear antiferromagnetism},\ }\href {http://link.aps.org/pdf/10.1103/PhysRevLett.126.127701} {\bibfield  {journal} {\bibinfo  {journal} {Phys. Rev. Lett.}\ }\textbf {\bibinfo {volume} {126}},\ \bibinfo {pages} {127701} (\bibinfo {year} {2021})}\BibitemShut {NoStop}%
    \bibitem [{\citenamefont {Zhou}\ \emph {et~al.}(2024)\citenamefont {Zhou}, \citenamefont {Feng}, \citenamefont {Zhang}, \citenamefont {Šmejkal}, \citenamefont {Sinova}, \citenamefont {Mokrousov},\ and\ \citenamefont {Yao}}]{Zhou2024-oa}%
      \BibitemOpen
      \bibfield  {author} {\bibinfo {author} {\bibfnamefont {X.}~\bibnamefont {Zhou}}, \bibinfo {author} {\bibfnamefont {W.}~\bibnamefont {Feng}}, \bibinfo {author} {\bibfnamefont {R.-W.}\ \bibnamefont {Zhang}}, \bibinfo {author} {\bibfnamefont {L.}~\bibnamefont {Šmejkal}}, \bibinfo {author} {\bibfnamefont {J.}~\bibnamefont {Sinova}}, \bibinfo {author} {\bibfnamefont {Y.}~\bibnamefont {Mokrousov}},\ and\ \bibinfo {author} {\bibfnamefont {Y.}~\bibnamefont {Yao}},\ }\bibfield  {title} {\bibinfo {title} {Crystal thermal transport in altermagnetic {RuO$_2$}},\ }\href {http://link.aps.org/pdf/10.1103/PhysRevLett.132.056701} {\bibfield  {journal} {\bibinfo  {journal} {Phys. Rev. Lett.}\ }\textbf {\bibinfo {volume} {132}},\ \bibinfo {pages} {056701} (\bibinfo {year} {2024})}\BibitemShut {NoStop}%
    \bibitem [{\citenamefont {Šmejkal}\ \emph {et~al.}(2022{\natexlab{c}})\citenamefont {Šmejkal}, \citenamefont {Hellenes}, \citenamefont {González-Hernández}, \citenamefont {Sinova},\ and\ \citenamefont {Jungwirth}}]{Smejkal2022-dz}%
      \BibitemOpen
      \bibfield  {author} {\bibinfo {author} {\bibfnamefont {L.}~\bibnamefont {Šmejkal}}, \bibinfo {author} {\bibfnamefont {A.~B.}\ \bibnamefont {Hellenes}}, \bibinfo {author} {\bibfnamefont {R.}~\bibnamefont {González-Hernández}}, \bibinfo {author} {\bibfnamefont {J.}~\bibnamefont {Sinova}},\ and\ \bibinfo {author} {\bibfnamefont {T.}~\bibnamefont {Jungwirth}},\ }\bibfield  {title} {\bibinfo {title} {Giant and tunneling magnetoresistance in unconventional collinear antiferromagnets with nonrelativistic spin-momentum coupling},\ }\href {http://link.aps.org/pdf/10.1103/PhysRevX.12.011028} {\bibfield  {journal} {\bibinfo  {journal} {Phys. Rev. X}\ }\textbf {\bibinfo {volume} {12}},\ \bibinfo {pages} {011028} (\bibinfo {year} {2022}{\natexlab{c}})}\BibitemShut {NoStop}%
    \bibitem [{\citenamefont {Tanaka}\ \emph {et~al.}(2024{\natexlab{a}})\citenamefont {Tanaka}, \citenamefont {Nomoto},\ and\ \citenamefont {Arita}}]{Tanaka2024-eu}%
      \BibitemOpen
      \bibfield  {author} {\bibinfo {author} {\bibfnamefont {K.}~\bibnamefont {Tanaka}}, \bibinfo {author} {\bibfnamefont {T.}~\bibnamefont {Nomoto}},\ and\ \bibinfo {author} {\bibfnamefont {R.}~\bibnamefont {Arita}},\ }\bibfield  {title} {\bibinfo {title} {First-principles study of the tunnel magnetoresistance effect with {Cr}-doped {RuO$_2$} electrode},\ }\href {https://journals.aps.org/prb/abstract/10.1103/PhysRevB.110.064433} {\bibfield  {journal} {\bibinfo  {journal} {Phys. Rev. B}\ }\textbf {\bibinfo {volume} {110}} (\bibinfo {year} {2024}{\natexlab{a}})}\BibitemShut {NoStop}%
    \bibitem [{\citenamefont {Tanaka}\ \emph {et~al.}(2024{\natexlab{b}})\citenamefont {Tanaka}, \citenamefont {Nomoto},\ and\ \citenamefont {Arita}}]{Tanaka2024-il}%
      \BibitemOpen
      \bibfield  {author} {\bibinfo {author} {\bibfnamefont {K.}~\bibnamefont {Tanaka}}, \bibinfo {author} {\bibfnamefont {T.}~\bibnamefont {Nomoto}},\ and\ \bibinfo {author} {\bibfnamefont {R.}~\bibnamefont {Arita}},\ }\bibfield  {title} {\bibinfo {title} {Approaches to tunnel magnetoresistance effect with antiferromagnets},\ }\href {http://arxiv.org/abs/2410.19513} {\bibfield  {journal} {\bibinfo  {journal} {arXiv [cond-mat.mes-hall]}\ } (\bibinfo {year} {2024}{\natexlab{b}})}\BibitemShut {NoStop}%
    \bibitem [{\citenamefont {Fernandes}\ \emph {et~al.}(2024)\citenamefont {Fernandes}, \citenamefont {de~Carvalho}, \citenamefont {Birol},\ and\ \citenamefont {Pereira}}]{Fernandes2024-db}%
      \BibitemOpen
      \bibfield  {author} {\bibinfo {author} {\bibfnamefont {R.~M.}\ \bibnamefont {Fernandes}}, \bibinfo {author} {\bibfnamefont {V.~S.}\ \bibnamefont {de~Carvalho}}, \bibinfo {author} {\bibfnamefont {T.}~\bibnamefont {Birol}},\ and\ \bibinfo {author} {\bibfnamefont {R.~G.}\ \bibnamefont {Pereira}},\ }\bibfield  {title} {\bibinfo {title} {Topological transition from nodal to nodeless zeeman splitting in altermagnets},\ }\href {https://journals.aps.org/prb/abstract/10.1103/PhysRevB.109.024404} {\bibfield  {journal} {\bibinfo  {journal} {Phys. Rev. B.}\ }\textbf {\bibinfo {volume} {109}} (\bibinfo {year} {2024})}\BibitemShut {NoStop}%
    \bibitem [{\citenamefont {Bhowal}\ and\ \citenamefont {Spaldin}(2024)}]{Bhowal2024-ty}%
      \BibitemOpen
      \bibfield  {author} {\bibinfo {author} {\bibfnamefont {S.}~\bibnamefont {Bhowal}}\ and\ \bibinfo {author} {\bibfnamefont {N.~A.}\ \bibnamefont {Spaldin}},\ }\bibfield  {title} {\bibinfo {title} {Ferroically ordered magnetic octupoles in d -wave altermagnets},\ }\href {https://journals.aps.org/prx/abstract/10.1103/PhysRevX.14.011019} {\bibfield  {journal} {\bibinfo  {journal} {Phys. Rev. X.}\ }\textbf {\bibinfo {volume} {14}} (\bibinfo {year} {2024})}\BibitemShut {NoStop}%
    \bibitem [{\citenamefont {Reimers}\ \emph {et~al.}(2024)\citenamefont {Reimers}, \citenamefont {Odenbreit}, \citenamefont {Šmejkal}, \citenamefont {Strocov}, \citenamefont {Constantinou}, \citenamefont {Hellenes}, \citenamefont {Jaeschke~Ubiergo}, \citenamefont {Campos}, \citenamefont {Bharadwaj}, \citenamefont {Chakraborty}, \citenamefont {Denneulin}, \citenamefont {Shi}, \citenamefont {Dunin-Borkowski}, \citenamefont {Das}, \citenamefont {Kläui}, \citenamefont {Sinova},\ and\ \citenamefont {Jourdan}}]{Reimers2024-tx}%
      \BibitemOpen
      \bibfield  {author} {\bibinfo {author} {\bibfnamefont {S.}~\bibnamefont {Reimers}}, \bibinfo {author} {\bibfnamefont {L.}~\bibnamefont {Odenbreit}}, \bibinfo {author} {\bibfnamefont {L.}~\bibnamefont {Šmejkal}}, \bibinfo {author} {\bibfnamefont {V.~N.}\ \bibnamefont {Strocov}}, \bibinfo {author} {\bibfnamefont {P.}~\bibnamefont {Constantinou}}, \bibinfo {author} {\bibfnamefont {A.~B.}\ \bibnamefont {Hellenes}}, \bibinfo {author} {\bibfnamefont {R.}~\bibnamefont {Jaeschke~Ubiergo}}, \bibinfo {author} {\bibfnamefont {W.~H.}\ \bibnamefont {Campos}}, \bibinfo {author} {\bibfnamefont {V.~K.}\ \bibnamefont {Bharadwaj}}, \bibinfo {author} {\bibfnamefont {A.}~\bibnamefont {Chakraborty}}, \bibinfo {author} {\bibfnamefont {T.}~\bibnamefont {Denneulin}}, \bibinfo {author} {\bibfnamefont {W.}~\bibnamefont {Shi}}, \bibinfo {author} {\bibfnamefont {R.~E.}\ \bibnamefont {Dunin-Borkowski}}, \bibinfo {author} {\bibfnamefont {S.}~\bibnamefont {Das}}, \bibinfo {author} {\bibfnamefont {M.}~\bibnamefont {Kläui}}, \bibinfo
      {author} {\bibfnamefont {J.}~\bibnamefont {Sinova}},\ and\ \bibinfo {author} {\bibfnamefont {M.}~\bibnamefont {Jourdan}},\ }\bibfield  {title} {\bibinfo {title} {Direct observation of altermagnetic band splitting in {CrSb} thin films},\ }\href {https://www.nature.com/articles/s41467-024-46476-5} {\bibfield  {journal} {\bibinfo  {journal} {Nat. Commun.}\ }\textbf {\bibinfo {volume} {15}},\ \bibinfo {pages} {2116} (\bibinfo {year} {2024})}\BibitemShut {NoStop}%
    \bibitem [{\citenamefont {Krempaský}\ \emph {et~al.}(2024)\citenamefont {Krempaský}, \citenamefont {Šmejkal}, \citenamefont {D'Souza}, \citenamefont {Hajlaoui}, \citenamefont {Springholz}, \citenamefont {Uhlířová}, \citenamefont {Alarab}, \citenamefont {Constantinou}, \citenamefont {Strocov}, \citenamefont {Usanov}, \citenamefont {Pudelko}, \citenamefont {González-Hernández}, \citenamefont {Birk~Hellenes}, \citenamefont {Jansa}, \citenamefont {Reichlová}, \citenamefont {Šobáň}, \citenamefont {Gonzalez~Betancourt}, \citenamefont {Wadley}, \citenamefont {Sinova}, \citenamefont {Kriegner}, \citenamefont {Minár}, \citenamefont {Dil},\ and\ \citenamefont {Jungwirth}}]{Krempasky2024-db}%
      \BibitemOpen
      \bibfield  {author} {\bibinfo {author} {\bibfnamefont {J.}~\bibnamefont {Krempaský}}, \bibinfo {author} {\bibfnamefont {L.}~\bibnamefont {Šmejkal}}, \bibinfo {author} {\bibfnamefont {S.~W.}\ \bibnamefont {D'Souza}}, \bibinfo {author} {\bibfnamefont {M.}~\bibnamefont {Hajlaoui}}, \bibinfo {author} {\bibfnamefont {G.}~\bibnamefont {Springholz}}, \bibinfo {author} {\bibfnamefont {K.}~\bibnamefont {Uhlířová}}, \bibinfo {author} {\bibfnamefont {F.}~\bibnamefont {Alarab}}, \bibinfo {author} {\bibfnamefont {P.~C.}\ \bibnamefont {Constantinou}}, \bibinfo {author} {\bibfnamefont {V.}~\bibnamefont {Strocov}}, \bibinfo {author} {\bibfnamefont {D.}~\bibnamefont {Usanov}}, \bibinfo {author} {\bibfnamefont {W.~R.}\ \bibnamefont {Pudelko}}, \bibinfo {author} {\bibfnamefont {R.}~\bibnamefont {González-Hernández}}, \bibinfo {author} {\bibfnamefont {A.}~\bibnamefont {Birk~Hellenes}}, \bibinfo {author} {\bibfnamefont {Z.}~\bibnamefont {Jansa}}, \bibinfo {author} {\bibfnamefont {H.}~\bibnamefont {Reichlová}}, \bibinfo
      {author} {\bibfnamefont {Z.}~\bibnamefont {Šobáň}}, \bibinfo {author} {\bibfnamefont {R.~D.}\ \bibnamefont {Gonzalez~Betancourt}}, \bibinfo {author} {\bibfnamefont {P.}~\bibnamefont {Wadley}}, \bibinfo {author} {\bibfnamefont {J.}~\bibnamefont {Sinova}}, \bibinfo {author} {\bibfnamefont {D.}~\bibnamefont {Kriegner}}, \bibinfo {author} {\bibfnamefont {J.}~\bibnamefont {Minár}}, \bibinfo {author} {\bibfnamefont {J.~H.}\ \bibnamefont {Dil}},\ and\ \bibinfo {author} {\bibfnamefont {T.}~\bibnamefont {Jungwirth}},\ }\bibfield  {title} {\bibinfo {title} {Altermagnetic lifting of {Kramers} spin degeneracy},\ }\href {https://www.nature.com/articles/s41586-023-06907-7} {\bibfield  {journal} {\bibinfo  {journal} {Nature}\ }\textbf {\bibinfo {volume} {626}},\ \bibinfo {pages} {517} (\bibinfo {year} {2024})}\BibitemShut {NoStop}%
    \bibitem [{\citenamefont {Takagi}\ \emph {et~al.}(2024)\citenamefont {Takagi}, \citenamefont {Hirakida}, \citenamefont {Settai}, \citenamefont {Oiwa}, \citenamefont {Takagi}, \citenamefont {Kitaori}, \citenamefont {Yamauchi}, \citenamefont {Inoue}, \citenamefont {Yamaura}, \citenamefont {Nishio-Hamane}, \citenamefont {Itoh}, \citenamefont {Aji}, \citenamefont {Saito}, \citenamefont {Nakajima}, \citenamefont {Nomoto}, \citenamefont {Arita},\ and\ \citenamefont {Seki}}]{Takagi2024-om}%
      \BibitemOpen
      \bibfield  {author} {\bibinfo {author} {\bibfnamefont {R.}~\bibnamefont {Takagi}}, \bibinfo {author} {\bibfnamefont {R.}~\bibnamefont {Hirakida}}, \bibinfo {author} {\bibfnamefont {Y.}~\bibnamefont {Settai}}, \bibinfo {author} {\bibfnamefont {R.}~\bibnamefont {Oiwa}}, \bibinfo {author} {\bibfnamefont {H.}~\bibnamefont {Takagi}}, \bibinfo {author} {\bibfnamefont {A.}~\bibnamefont {Kitaori}}, \bibinfo {author} {\bibfnamefont {K.}~\bibnamefont {Yamauchi}}, \bibinfo {author} {\bibfnamefont {H.}~\bibnamefont {Inoue}}, \bibinfo {author} {\bibfnamefont {J.-I.}\ \bibnamefont {Yamaura}}, \bibinfo {author} {\bibfnamefont {D.}~\bibnamefont {Nishio-Hamane}}, \bibinfo {author} {\bibfnamefont {S.}~\bibnamefont {Itoh}}, \bibinfo {author} {\bibfnamefont {S.}~\bibnamefont {Aji}}, \bibinfo {author} {\bibfnamefont {H.}~\bibnamefont {Saito}}, \bibinfo {author} {\bibfnamefont {T.}~\bibnamefont {Nakajima}}, \bibinfo {author} {\bibfnamefont {T.}~\bibnamefont {Nomoto}}, \bibinfo {author} {\bibfnamefont {R.}~\bibnamefont {Arita}},\
      and\ \bibinfo {author} {\bibfnamefont {S.}~\bibnamefont {Seki}},\ }\bibfield  {title} {\bibinfo {title} {Spontaneous hall effect induced by collinear antiferromagnetic order at room temperature},\ }\href {https://www.nature.com/articles/s41563-024-02058-w} {\bibfield  {journal} {\bibinfo  {journal} {Nat. Mater.}\ ,\ \bibinfo {pages} {1}} (\bibinfo {year} {2024})}\BibitemShut {NoStop}%
    \bibitem [{\citenamefont {Reichlova}\ \emph {et~al.}(2024)\citenamefont {Reichlova}, \citenamefont {Lopes~Seeger}, \citenamefont {González-Hernández}, \citenamefont {Kounta}, \citenamefont {Schlitz}, \citenamefont {Kriegner}, \citenamefont {Ritzinger}, \citenamefont {Lammel}, \citenamefont {Leiviskä}, \citenamefont {Birk~Hellenes}, \citenamefont {Olejník}, \citenamefont {Petřiček}, \citenamefont {Doležal}, \citenamefont {Horak}, \citenamefont {Schmoranzerova}, \citenamefont {Badura}, \citenamefont {Bertaina}, \citenamefont {Thomas}, \citenamefont {Baltz}, \citenamefont {Michez}, \citenamefont {Sinova}, \citenamefont {Goennenwein}, \citenamefont {Jungwirth},\ and\ \citenamefont {Šmejkal}}]{Reichlova2024-fo}%
      \BibitemOpen
      \bibfield  {author} {\bibinfo {author} {\bibfnamefont {H.}~\bibnamefont {Reichlova}}, \bibinfo {author} {\bibfnamefont {R.}~\bibnamefont {Lopes~Seeger}}, \bibinfo {author} {\bibfnamefont {R.}~\bibnamefont {González-Hernández}}, \bibinfo {author} {\bibfnamefont {I.}~\bibnamefont {Kounta}}, \bibinfo {author} {\bibfnamefont {R.}~\bibnamefont {Schlitz}}, \bibinfo {author} {\bibfnamefont {D.}~\bibnamefont {Kriegner}}, \bibinfo {author} {\bibfnamefont {P.}~\bibnamefont {Ritzinger}}, \bibinfo {author} {\bibfnamefont {M.}~\bibnamefont {Lammel}}, \bibinfo {author} {\bibfnamefont {M.}~\bibnamefont {Leiviskä}}, \bibinfo {author} {\bibfnamefont {A.}~\bibnamefont {Birk~Hellenes}}, \bibinfo {author} {\bibfnamefont {K.}~\bibnamefont {Olejník}}, \bibinfo {author} {\bibfnamefont {V.}~\bibnamefont {Petřiček}}, \bibinfo {author} {\bibfnamefont {P.}~\bibnamefont {Doležal}}, \bibinfo {author} {\bibfnamefont {L.}~\bibnamefont {Horak}}, \bibinfo {author} {\bibfnamefont {E.}~\bibnamefont {Schmoranzerova}}, \bibinfo {author}
      {\bibfnamefont {A.}~\bibnamefont {Badura}}, \bibinfo {author} {\bibfnamefont {S.}~\bibnamefont {Bertaina}}, \bibinfo {author} {\bibfnamefont {A.}~\bibnamefont {Thomas}}, \bibinfo {author} {\bibfnamefont {V.}~\bibnamefont {Baltz}}, \bibinfo {author} {\bibfnamefont {L.}~\bibnamefont {Michez}}, \bibinfo {author} {\bibfnamefont {J.}~\bibnamefont {Sinova}}, \bibinfo {author} {\bibfnamefont {S.~T.~B.}\ \bibnamefont {Goennenwein}}, \bibinfo {author} {\bibfnamefont {T.}~\bibnamefont {Jungwirth}},\ and\ \bibinfo {author} {\bibfnamefont {L.}~\bibnamefont {Šmejkal}},\ }\bibfield  {title} {\bibinfo {title} {Observation of a spontaneous anomalous hall response in the {Mn$_5$Si$_3$} $d$-wave altermagnet candidate},\ }\href {https://www.nature.com/articles/s41467-024-48493-w} {\bibfield  {journal} {\bibinfo  {journal} {Nat. Commun.}\ }\textbf {\bibinfo {volume} {15}},\ \bibinfo {pages} {4961} (\bibinfo {year} {2024})}\BibitemShut {NoStop}%
    \bibitem [{\citenamefont {Osumi}\ \emph {et~al.}(2024)\citenamefont {Osumi}, \citenamefont {Souma}, \citenamefont {Aoyama}, \citenamefont {Yamauchi}, \citenamefont {Honma}, \citenamefont {Nakayama}, \citenamefont {Takahashi}, \citenamefont {Ohgushi},\ and\ \citenamefont {Sato}}]{Osumi2024-lb}%
      \BibitemOpen
      \bibfield  {author} {\bibinfo {author} {\bibfnamefont {T.}~\bibnamefont {Osumi}}, \bibinfo {author} {\bibfnamefont {S.}~\bibnamefont {Souma}}, \bibinfo {author} {\bibfnamefont {T.}~\bibnamefont {Aoyama}}, \bibinfo {author} {\bibfnamefont {K.}~\bibnamefont {Yamauchi}}, \bibinfo {author} {\bibfnamefont {A.}~\bibnamefont {Honma}}, \bibinfo {author} {\bibfnamefont {K.}~\bibnamefont {Nakayama}}, \bibinfo {author} {\bibfnamefont {T.}~\bibnamefont {Takahashi}}, \bibinfo {author} {\bibfnamefont {K.}~\bibnamefont {Ohgushi}},\ and\ \bibinfo {author} {\bibfnamefont {T.}~\bibnamefont {Sato}},\ }\bibfield  {title} {\bibinfo {title} {Observation of a giant band splitting in altermagnetic {MnTe}},\ }\href {http://link.aps.org/pdf/10.1103/PhysRevB.109.115102} {\bibfield  {journal} {\bibinfo  {journal} {Phys. Rev. B.}\ }\textbf {\bibinfo {volume} {109}},\ \bibinfo {pages} {115102} (\bibinfo {year} {2024})}\BibitemShut {NoStop}%
    \bibitem [{\citenamefont {Lee}\ \emph {et~al.}(2024)\citenamefont {Lee}, \citenamefont {Lee}, \citenamefont {Jung}, \citenamefont {Jung}, \citenamefont {Kim}, \citenamefont {Lee}, \citenamefont {Seok}, \citenamefont {Kim}, \citenamefont {Park}, \citenamefont {Šmejkal}, \citenamefont {Kang},\ and\ \citenamefont {Kim}}]{Lee2024-jk}%
      \BibitemOpen
      \bibfield  {author} {\bibinfo {author} {\bibfnamefont {S.}~\bibnamefont {Lee}}, \bibinfo {author} {\bibfnamefont {S.}~\bibnamefont {Lee}}, \bibinfo {author} {\bibfnamefont {S.}~\bibnamefont {Jung}}, \bibinfo {author} {\bibfnamefont {J.}~\bibnamefont {Jung}}, \bibinfo {author} {\bibfnamefont {D.}~\bibnamefont {Kim}}, \bibinfo {author} {\bibfnamefont {Y.}~\bibnamefont {Lee}}, \bibinfo {author} {\bibfnamefont {B.}~\bibnamefont {Seok}}, \bibinfo {author} {\bibfnamefont {J.}~\bibnamefont {Kim}}, \bibinfo {author} {\bibfnamefont {B.~G.}\ \bibnamefont {Park}}, \bibinfo {author} {\bibfnamefont {L.}~\bibnamefont {Šmejkal}}, \bibinfo {author} {\bibfnamefont {C.-J.}\ \bibnamefont {Kang}},\ and\ \bibinfo {author} {\bibfnamefont {C.}~\bibnamefont {Kim}},\ }\bibfield  {title} {\bibinfo {title} {Broken kramers degeneracy in altermagnetic {MnTe}},\ }\href {https://journals.aps.org/prl/abstract/10.1103/PhysRevLett.132.036702} {\bibfield  {journal} {\bibinfo  {journal} {Phys. Rev. Lett.}\ }\textbf {\bibinfo {volume}
      {132}},\ \bibinfo {pages} {036702} (\bibinfo {year} {2024})}\BibitemShut {NoStop}%
    \bibitem [{Note1()}]{Note1}%
      \BibitemOpen
      \bibinfo {note} {Let us write the spin-momentum-locking texture of the electronic bands in collinear antiferromagnets as $F(\protect \bm {k})S_z$ with the form factor $F (\protect \bm {k})$ and $z$-component of spin $S_z$. Applying the $[-1 \cdot C_2 ||1]$ symmetry ($-1$ in the spin space means the $\protect \mathcal {T}${} operation) to $F(\protect \bm {k})S_z$ yields $F(\protect \bm {k}) = F(-\protect \bm {k})$~\cite {Smejkal2022-uk}.}\BibitemShut {Stop}%
    \bibitem [{\citenamefont {Sivianes}\ \emph {et~al.}(2024)\citenamefont {Sivianes}, \citenamefont {Santos},\ and\ \citenamefont {Ibañez-Azpiroz}}]{Sivianes2024-hk}%
      \BibitemOpen
      \bibfield  {author} {\bibinfo {author} {\bibfnamefont {J.}~\bibnamefont {Sivianes}}, \bibinfo {author} {\bibfnamefont {F.~J.~d.}\ \bibnamefont {Santos}},\ and\ \bibinfo {author} {\bibfnamefont {J.}~\bibnamefont {Ibañez-Azpiroz}},\ }\bibfield  {title} {\bibinfo {title} {Unconventional $p$-wave magnets as sources of nonlinear photocurrents},\ }\href {http://arxiv.org/abs/2406.19842} {\bibfield  {journal} {\bibinfo  {journal} {arXiv [cond-mat.mes-hall]}\ } (\bibinfo {year} {2024})}\BibitemShut {NoStop}%
    \bibitem [{\citenamefont {Brekke}\ \emph {et~al.}(2024)\citenamefont {Brekke}, \citenamefont {Sukhachov}, \citenamefont {Giil}, \citenamefont {Brataas},\ and\ \citenamefont {Linder}}]{Brekke2024-zk}%
      \BibitemOpen
      \bibfield  {author} {\bibinfo {author} {\bibfnamefont {B.}~\bibnamefont {Brekke}}, \bibinfo {author} {\bibfnamefont {P.}~\bibnamefont {Sukhachov}}, \bibinfo {author} {\bibfnamefont {H.~G.}\ \bibnamefont {Giil}}, \bibinfo {author} {\bibfnamefont {A.}~\bibnamefont {Brataas}},\ and\ \bibinfo {author} {\bibfnamefont {J.}~\bibnamefont {Linder}},\ }\bibfield  {title} {\bibinfo {title} {Minimal models and transport properties of unconventional $p$-wave magnets},\ }\href {http://arxiv.org/abs/2405.15823} {\bibfield  {journal} {\bibinfo  {journal} {arXiv [cond-mat.mes-hall]}\ } (\bibinfo {year} {2024})}\BibitemShut {NoStop}%
    \bibitem [{\citenamefont {Hellenes}\ \emph {et~al.}(2023)\citenamefont {Hellenes}, \citenamefont {Jungwirth}, \citenamefont {Jaeschke-Ubiergo}, \citenamefont {Chakraborty}, \citenamefont {Sinova},\ and\ \citenamefont {{\v{S}}mejkal}}]{Hellenes2023-ro}%
      \BibitemOpen
      \bibfield  {author} {\bibinfo {author} {\bibfnamefont {A.~B.}\ \bibnamefont {Hellenes}}, \bibinfo {author} {\bibfnamefont {T.}~\bibnamefont {Jungwirth}}, \bibinfo {author} {\bibfnamefont {R.}~\bibnamefont {Jaeschke-Ubiergo}}, \bibinfo {author} {\bibfnamefont {A.}~\bibnamefont {Chakraborty}}, \bibinfo {author} {\bibfnamefont {J.}~\bibnamefont {Sinova}},\ and\ \bibinfo {author} {\bibfnamefont {L.}~\bibnamefont {{\v{S}}mejkal}},\ }\bibfield  {title} {\bibinfo {title} {{P}-wave magnets},\ }\href {http://arxiv.org/abs/2309.01607} {\bibfield  {journal} {\bibinfo  {journal} {arXiv [cond-mat.mes-hall]}\ } (\bibinfo {year} {2023})}\BibitemShut {NoStop}%
    \bibitem [{\citenamefont {Pari}\ \emph {et~al.}(2024)\citenamefont {Pari}, \citenamefont {Jaeschke-Ubiergo}, \citenamefont {Chakraborty}, \citenamefont {{\v{S}}mejkal},\ and\ \citenamefont {Sinova}}]{Pari2024-mv}%
      \BibitemOpen
      \bibfield  {author} {\bibinfo {author} {\bibfnamefont {N.~A.~{\'{A}}.}\ \bibnamefont {Pari}}, \bibinfo {author} {\bibfnamefont {R.}~\bibnamefont {Jaeschke-Ubiergo}}, \bibinfo {author} {\bibfnamefont {A.}~\bibnamefont {Chakraborty}}, \bibinfo {author} {\bibfnamefont {L.}~\bibnamefont {{\v{S}}mejkal}},\ and\ \bibinfo {author} {\bibfnamefont {J.}~\bibnamefont {Sinova}},\ }\bibfield  {title} {\bibinfo {title} {Non-relativistic linear edelstein effect in non-collinear {EuIn$_2$As$_2$}},\ }\href {http://arxiv.org/abs/2412.10984} {\bibfield  {journal} {\bibinfo  {journal} {arXiv [cond-mat.str-el]}\ } (\bibinfo {year} {2024})}\BibitemShut {NoStop}%
    \bibitem [{\citenamefont {Chakraborty}\ \emph {et~al.}(2024)\citenamefont {Chakraborty}, \citenamefont {Hellenes}, \citenamefont {Jaeschke-Ubiergo}, \citenamefont {Jungwirth}, \citenamefont {{\v{S}}mejkal},\ and\ \citenamefont {Sinova}}]{Chakraborty2024-xy}%
      \BibitemOpen
      \bibfield  {author} {\bibinfo {author} {\bibfnamefont {A.}~\bibnamefont {Chakraborty}}, \bibinfo {author} {\bibfnamefont {A.~B.}\ \bibnamefont {Hellenes}}, \bibinfo {author} {\bibfnamefont {R.}~\bibnamefont {Jaeschke-Ubiergo}}, \bibinfo {author} {\bibfnamefont {T.}~\bibnamefont {Jungwirth}}, \bibinfo {author} {\bibfnamefont {L.}~\bibnamefont {{\v{S}}mejkal}},\ and\ \bibinfo {author} {\bibfnamefont {J.}~\bibnamefont {Sinova}},\ }\bibfield  {title} {\bibinfo {title} {Highly efficient non-relativistic edelstein effect in p-wave magnets},\ }\href {http://arxiv.org/abs/2411.16378} {\bibfield  {journal} {\bibinfo  {journal} {arXiv [cond-mat.mes-hall]}\ } (\bibinfo {year} {2024})}\BibitemShut {NoStop}%
    \bibitem [{\citenamefont {Watanabe}\ and\ \citenamefont {Yanase}(2024)}]{Watanabe2024-qo}%
      \BibitemOpen
      \bibfield  {author} {\bibinfo {author} {\bibfnamefont {H.}~\bibnamefont {Watanabe}}\ and\ \bibinfo {author} {\bibfnamefont {Y.}~\bibnamefont {Yanase}},\ }\bibfield  {title} {\bibinfo {title} {Magnetic parity violation and parity-time-reversal-symmetric magnets},\ }\href {https://iopscience.iop.org/article/10.1088/1361-648X/ad52dd/meta} {\bibfield  {journal} {\bibinfo  {journal} {J. Phys. Condens. Matter}\ }\textbf {\bibinfo {volume} {36}},\ \bibinfo {pages} {373001} (\bibinfo {year} {2024})}\BibitemShut {NoStop}%
    \bibitem [{\citenamefont {Dzyaloshinskii}(1959)}]{Dzyaloshinskii1959-gm}%
      \BibitemOpen
      \bibfield  {author} {\bibinfo {author} {\bibfnamefont {I.~E.}\ \bibnamefont {Dzyaloshinskii}},\ }\bibfield  {title} {\bibinfo {title} {On the magneto-electrical effect in antiferromagnets},\ }\href@noop {} {\bibfield  {journal} {\bibinfo  {journal} {J. Exp. Theor. Phys}\ }\textbf {\bibinfo {volume} {37}},\ \bibinfo {pages} {881} (\bibinfo {year} {1959})},\ \bibinfo {note} {[I.~E.~Dzyaloshinskii, Zh. Exp. Teor. Fiz. \textbf{37}, 881 (1959)]}\BibitemShut {NoStop}%
    \bibitem [{\citenamefont {Astrov}(1960)}]{Astrov1960-kw}%
      \BibitemOpen
      \bibfield  {author} {\bibinfo {author} {\bibfnamefont {D.~N.}\ \bibnamefont {Astrov}},\ }\bibfield  {title} {\bibinfo {title} {The magnetoelectric effect in antiferromagnetics},\ }\href {http://www.jetp.ras.ru/cgi-bin/dn/e_011_03_0708.pdf} {\bibfield  {journal} {\bibinfo  {journal} {Sov. Phys. JETP}\ }\textbf {\bibinfo {volume} {11}},\ \bibinfo {pages} {708} (\bibinfo {year} {1960})},\ \bibinfo {note} {[D.~N.~Astrov, Zh. Exp. Teor. Fiz. \textbf{38}, 984 (1960)].}\BibitemShut {Stop}%
    \bibitem [{\citenamefont {Rado}\ and\ \citenamefont {Folen}(1961)}]{Rado1961-tg}%
      \BibitemOpen
      \bibfield  {author} {\bibinfo {author} {\bibfnamefont {G.~T.}\ \bibnamefont {Rado}}\ and\ \bibinfo {author} {\bibfnamefont {V.~J.}\ \bibnamefont {Folen}},\ }\bibfield  {title} {\bibinfo {title} {Observation of the magnetically induced magnetoelectric effect and evidence for antiferromagnetic domains},\ }\href {https://journals.aps.org/prl/abstract/10.1103/PhysRevLett.7.310} {\bibfield  {journal} {\bibinfo  {journal} {Phys. Rev. Lett.}\ }\textbf {\bibinfo {volume} {7}},\ \bibinfo {pages} {310} (\bibinfo {year} {1961})}\BibitemShut {NoStop}%
    \bibitem [{\citenamefont {Wadley}\ \emph {et~al.}(2016)\citenamefont {Wadley}, \citenamefont {Howells}, \citenamefont {{\v{Z}}elezn{\'{y}}}, \citenamefont {Andrews}, \citenamefont {Hills}, \citenamefont {Campion}, \citenamefont {Novák}, \citenamefont {Olejník}, \citenamefont {Maccherozzi}, \citenamefont {Dhesi}, \citenamefont {Martin}, \citenamefont {Wagner}, \citenamefont {Wunderlich}, \citenamefont {Freimuth}, \citenamefont {Mokrousov}, \citenamefont {Kune{\v{s}}}, \citenamefont {Chauhan}, \citenamefont {Grzybowski}, \citenamefont {Rushforth}, \citenamefont {Edmonds}, \citenamefont {Gallagher},\ and\ \citenamefont {Jungwirth}}]{Wadley2016-rx}%
      \BibitemOpen
      \bibfield  {author} {\bibinfo {author} {\bibfnamefont {P.}~\bibnamefont {Wadley}}, \bibinfo {author} {\bibfnamefont {B.}~\bibnamefont {Howells}}, \bibinfo {author} {\bibfnamefont {J.}~\bibnamefont {{\v{Z}}elezn{\'{y}}}}, \bibinfo {author} {\bibfnamefont {C.}~\bibnamefont {Andrews}}, \bibinfo {author} {\bibfnamefont {V.}~\bibnamefont {Hills}}, \bibinfo {author} {\bibfnamefont {R.~P.}\ \bibnamefont {Campion}}, \bibinfo {author} {\bibfnamefont {V.}~\bibnamefont {Novák}}, \bibinfo {author} {\bibfnamefont {K.}~\bibnamefont {Olejník}}, \bibinfo {author} {\bibfnamefont {F.}~\bibnamefont {Maccherozzi}}, \bibinfo {author} {\bibfnamefont {S.~S.}\ \bibnamefont {Dhesi}}, \bibinfo {author} {\bibfnamefont {S.~Y.}\ \bibnamefont {Martin}}, \bibinfo {author} {\bibfnamefont {T.}~\bibnamefont {Wagner}}, \bibinfo {author} {\bibfnamefont {J.}~\bibnamefont {Wunderlich}}, \bibinfo {author} {\bibfnamefont {F.}~\bibnamefont {Freimuth}}, \bibinfo {author} {\bibfnamefont {Y.}~\bibnamefont {Mokrousov}}, \bibinfo {author}
      {\bibfnamefont {J.}~\bibnamefont {Kune{\v{s}}}}, \bibinfo {author} {\bibfnamefont {J.~S.}\ \bibnamefont {Chauhan}}, \bibinfo {author} {\bibfnamefont {M.~J.}\ \bibnamefont {Grzybowski}}, \bibinfo {author} {\bibfnamefont {A.~W.}\ \bibnamefont {Rushforth}}, \bibinfo {author} {\bibfnamefont {K.~W.}\ \bibnamefont {Edmonds}}, \bibinfo {author} {\bibfnamefont {B.~L.}\ \bibnamefont {Gallagher}},\ and\ \bibinfo {author} {\bibfnamefont {T.}~\bibnamefont {Jungwirth}},\ }\bibfield  {title} {\bibinfo {title} {Electrical switching of an antiferromagnet},\ }\href {https://www.science.org/doi/10.1126/science.aab1031} {\bibfield  {journal} {\bibinfo  {journal} {Science}\ }\textbf {\bibinfo {volume} {351}},\ \bibinfo {pages} {587} (\bibinfo {year} {2016})}\BibitemShut {NoStop}%
    \bibitem [{\citenamefont {Bodnar}\ \emph {et~al.}(2018)\citenamefont {Bodnar}, \citenamefont {Šmejkal}, \citenamefont {Turek}, \citenamefont {Jungwirth}, \citenamefont {Gomonay}, \citenamefont {Sinova}, \citenamefont {Sapozhnik}, \citenamefont {Elmers}, \citenamefont {Kläui},\ and\ \citenamefont {Jourdan}}]{Bodnar2018-jg}%
      \BibitemOpen
      \bibfield  {author} {\bibinfo {author} {\bibfnamefont {S.~Y.}\ \bibnamefont {Bodnar}}, \bibinfo {author} {\bibfnamefont {L.}~\bibnamefont {Šmejkal}}, \bibinfo {author} {\bibfnamefont {I.}~\bibnamefont {Turek}}, \bibinfo {author} {\bibfnamefont {T.}~\bibnamefont {Jungwirth}}, \bibinfo {author} {\bibfnamefont {O.}~\bibnamefont {Gomonay}}, \bibinfo {author} {\bibfnamefont {J.}~\bibnamefont {Sinova}}, \bibinfo {author} {\bibfnamefont {A.~A.}\ \bibnamefont {Sapozhnik}}, \bibinfo {author} {\bibfnamefont {H.-J.}\ \bibnamefont {Elmers}}, \bibinfo {author} {\bibfnamefont {M.}~\bibnamefont {Kläui}},\ and\ \bibinfo {author} {\bibfnamefont {M.}~\bibnamefont {Jourdan}},\ }\bibfield  {title} {\bibinfo {title} {Writing and reading antiferromagnetic {Mn$_2$Au} by néel spin-orbit torques and large anisotropic magnetoresistance},\ }\href {https://www.nature.com/articles/s41467-017-02780-x} {\bibfield  {journal} {\bibinfo  {journal} {Nat. Commun.}\ }\textbf {\bibinfo {volume} {9}},\ \bibinfo {pages} {348} (\bibinfo {year}
      {2018})}\BibitemShut {NoStop}%
    \bibitem [{\citenamefont {Yuan}\ \emph {et~al.}(2023)\citenamefont {Yuan}, \citenamefont {Zhang}, \citenamefont {Acosta},\ and\ \citenamefont {Zunger}}]{Yuan2023-wv}%
      \BibitemOpen
      \bibfield  {author} {\bibinfo {author} {\bibfnamefont {L.-D.}\ \bibnamefont {Yuan}}, \bibinfo {author} {\bibfnamefont {X.}~\bibnamefont {Zhang}}, \bibinfo {author} {\bibfnamefont {C.~M.}\ \bibnamefont {Acosta}},\ and\ \bibinfo {author} {\bibfnamefont {A.}~\bibnamefont {Zunger}},\ }\bibfield  {title} {\bibinfo {title} {Uncovering spin-orbit coupling-independent hidden spin polarization of energy bands in antiferromagnets},\ }\href {http://dx.doi.org/10.1038/s41467-023-40877-8} {\bibfield  {journal} {\bibinfo  {journal} {Nat. Commun.}\ }\textbf {\bibinfo {volume} {14}},\ \bibinfo {pages} {5301} (\bibinfo {year} {2023})}\BibitemShut {NoStop}%
    \bibitem [{\citenamefont {Guo}(2024)}]{Guo2024-pl}%
      \BibitemOpen
      \bibfield  {author} {\bibinfo {author} {\bibfnamefont {S.-D.}\ \bibnamefont {Guo}},\ }\bibfield  {title} {\bibinfo {title} {Hidden altermagnetism},\ }\href {http://arxiv.org/abs/2411.13795} {\bibfield  {journal} {\bibinfo  {journal} {arXiv [cond-mat.mtrl-sci]}\ } (\bibinfo {year} {2024})}\BibitemShut {NoStop}%
    \bibitem [{\citenamefont {Litvin}\ and\ \citenamefont {Opechowski}(1974)}]{Litvin1974-qv}%
      \BibitemOpen
      \bibfield  {author} {\bibinfo {author} {\bibfnamefont {D.~B.}\ \bibnamefont {Litvin}}\ and\ \bibinfo {author} {\bibfnamefont {W.}~\bibnamefont {Opechowski}},\ }\bibfield  {title} {\bibinfo {title} {Spin groups},\ }\href {http://dx.doi.org/10.1016/0031-8914(74)90157-8} {\bibfield  {journal} {\bibinfo  {journal} {Physica}\ }\textbf {\bibinfo {volume} {76}},\ \bibinfo {pages} {538} (\bibinfo {year} {1974})}\BibitemShut {NoStop}%
    \bibitem [{\citenamefont {Litvin}(1977)}]{Litvin1977-ti}%
      \BibitemOpen
      \bibfield  {author} {\bibinfo {author} {\bibfnamefont {D.~B.}\ \bibnamefont {Litvin}},\ }\bibfield  {title} {\bibinfo {title} {Spin point groups},\ }\href {https://journals.iucr.org/paper?a14103} {\bibfield  {journal} {\bibinfo  {journal} {Acta Crystallographica Section A: Crystal Physics, Diffraction, Theoretical and General Crystallography}\ }\textbf {\bibinfo {volume} {33}},\ \bibinfo {pages} {279} (\bibinfo {year} {1977})}\BibitemShut {NoStop}%
    \bibitem [{\citenamefont {Brinkman}\ and\ \citenamefont {Elliott}(1966)}]{Brinkman1966-si}%
      \BibitemOpen
      \bibfield  {author} {\bibinfo {author} {\bibfnamefont {W.~F.}\ \bibnamefont {Brinkman}}\ and\ \bibinfo {author} {\bibfnamefont {R.~J.}\ \bibnamefont {Elliott}},\ }\bibfield  {title} {\bibinfo {title} {Theory of spin-space groups},\ }\href {https://royalsocietypublishing.org/doi/10.1098/rspa.1966.0211} {\bibfield  {journal} {\bibinfo  {journal} {Proc. R. Soc. Lond.}\ }\textbf {\bibinfo {volume} {294}},\ \bibinfo {pages} {343} (\bibinfo {year} {1966})}\BibitemShut {NoStop}%
    \bibitem [{\citenamefont {Jiang}\ \emph {et~al.}(2024)\citenamefont {Jiang}, \citenamefont {Song}, \citenamefont {Zhu}, \citenamefont {Fang}, \citenamefont {Weng}, \citenamefont {Liu}, \citenamefont {Yang},\ and\ \citenamefont {Fang}}]{Jiang2024-oz}%
      \BibitemOpen
      \bibfield  {author} {\bibinfo {author} {\bibfnamefont {Y.}~\bibnamefont {Jiang}}, \bibinfo {author} {\bibfnamefont {Z.}~\bibnamefont {Song}}, \bibinfo {author} {\bibfnamefont {T.}~\bibnamefont {Zhu}}, \bibinfo {author} {\bibfnamefont {Z.}~\bibnamefont {Fang}}, \bibinfo {author} {\bibfnamefont {H.}~\bibnamefont {Weng}}, \bibinfo {author} {\bibfnamefont {Z.-X.}\ \bibnamefont {Liu}}, \bibinfo {author} {\bibfnamefont {J.}~\bibnamefont {Yang}},\ and\ \bibinfo {author} {\bibfnamefont {C.}~\bibnamefont {Fang}},\ }\bibfield  {title} {\bibinfo {title} {Enumeration of spin-space groups: {Toward} a complete description of symmetries of magnetic orders},\ }\href {http://link.aps.org/pdf/10.1103/PhysRevX.14.031039} {\bibfield  {journal} {\bibinfo  {journal} {Phys. Rev. X}\ }\textbf {\bibinfo {volume} {14}},\ \bibinfo {pages} {031039} (\bibinfo {year} {2024})}\BibitemShut {NoStop}%
    \bibitem [{\citenamefont {Chen}\ \emph {et~al.}(2024)\citenamefont {Chen}, \citenamefont {Ren}, \citenamefont {Zhu}, \citenamefont {Yu}, \citenamefont {Zhang}, \citenamefont {Liu}, \citenamefont {Li}, \citenamefont {Liu}, \citenamefont {Li},\ and\ \citenamefont {Liu}}]{Chen2024-td}%
      \BibitemOpen
      \bibfield  {author} {\bibinfo {author} {\bibfnamefont {X.}~\bibnamefont {Chen}}, \bibinfo {author} {\bibfnamefont {J.}~\bibnamefont {Ren}}, \bibinfo {author} {\bibfnamefont {Y.}~\bibnamefont {Zhu}}, \bibinfo {author} {\bibfnamefont {Y.}~\bibnamefont {Yu}}, \bibinfo {author} {\bibfnamefont {A.}~\bibnamefont {Zhang}}, \bibinfo {author} {\bibfnamefont {P.}~\bibnamefont {Liu}}, \bibinfo {author} {\bibfnamefont {J.}~\bibnamefont {Li}}, \bibinfo {author} {\bibfnamefont {Y.}~\bibnamefont {Liu}}, \bibinfo {author} {\bibfnamefont {C.}~\bibnamefont {Li}},\ and\ \bibinfo {author} {\bibfnamefont {Q.}~\bibnamefont {Liu}},\ }\bibfield  {title} {\bibinfo {title} {Enumeration and representation theory of spin space groups},\ }\href {http://link.aps.org/pdf/10.1103/PhysRevX.14.031038} {\bibfield  {journal} {\bibinfo  {journal} {Phys. Rev. X}\ }\textbf {\bibinfo {volume} {14}},\ \bibinfo {pages} {031038} (\bibinfo {year} {2024})}\BibitemShut {NoStop}%
    \bibitem [{\citenamefont {Xiao}\ \emph {et~al.}(2024)\citenamefont {Xiao}, \citenamefont {Zhao}, \citenamefont {Li}, \citenamefont {Shindou},\ and\ \citenamefont {Song}}]{Xiao2024-im}%
      \BibitemOpen
      \bibfield  {author} {\bibinfo {author} {\bibfnamefont {Z.}~\bibnamefont {Xiao}}, \bibinfo {author} {\bibfnamefont {J.}~\bibnamefont {Zhao}}, \bibinfo {author} {\bibfnamefont {Y.}~\bibnamefont {Li}}, \bibinfo {author} {\bibfnamefont {R.}~\bibnamefont {Shindou}},\ and\ \bibinfo {author} {\bibfnamefont {Z.-D.}\ \bibnamefont {Song}},\ }\bibfield  {title} {\bibinfo {title} {Spin space groups: {Full} classification and applications},\ }\href {https://journals.aps.org/prx/abstract/10.1103/PhysRevX.14.031037} {\bibfield  {journal} {\bibinfo  {journal} {Phys. Rev. X}\ }\textbf {\bibinfo {volume} {14}} (\bibinfo {year} {2024})}\BibitemShut {NoStop}%
    \bibitem [{\citenamefont {Watanabe}\ \emph {et~al.}(2024)\citenamefont {Watanabe}, \citenamefont {Shinohara}, \citenamefont {Nomoto}, \citenamefont {Togo},\ and\ \citenamefont {Arita}}]{Watanabe2024-mk}%
      \BibitemOpen
      \bibfield  {author} {\bibinfo {author} {\bibfnamefont {H.}~\bibnamefont {Watanabe}}, \bibinfo {author} {\bibfnamefont {K.}~\bibnamefont {Shinohara}}, \bibinfo {author} {\bibfnamefont {T.}~\bibnamefont {Nomoto}}, \bibinfo {author} {\bibfnamefont {A.}~\bibnamefont {Togo}},\ and\ \bibinfo {author} {\bibfnamefont {R.}~\bibnamefont {Arita}},\ }\bibfield  {title} {\bibinfo {title} {Symmetry analysis with spin crystallographic groups: Disentangling effects free of spin-orbit coupling in emergent electromagnetism},\ }\href {https://link.aps.org/doi/10.1103/PhysRevB.109.094438} {\bibfield  {journal} {\bibinfo  {journal} {Phys. Rev. B}\ }\textbf {\bibinfo {volume} {109}},\ \bibinfo {pages} {094438} (\bibinfo {year} {2024})}\BibitemShut {NoStop}%
    \bibitem [{\citenamefont {Shinohara}\ \emph {et~al.}(2024)\citenamefont {Shinohara}, \citenamefont {Togo}, \citenamefont {Watanabe}, \citenamefont {Nomoto}, \citenamefont {Tanaka},\ and\ \citenamefont {Arita}}]{Shinohara2024-pe}%
      \BibitemOpen
      \bibfield  {author} {\bibinfo {author} {\bibfnamefont {K.}~\bibnamefont {Shinohara}}, \bibinfo {author} {\bibfnamefont {A.}~\bibnamefont {Togo}}, \bibinfo {author} {\bibfnamefont {H.}~\bibnamefont {Watanabe}}, \bibinfo {author} {\bibfnamefont {T.}~\bibnamefont {Nomoto}}, \bibinfo {author} {\bibfnamefont {I.}~\bibnamefont {Tanaka}},\ and\ \bibinfo {author} {\bibfnamefont {R.}~\bibnamefont {Arita}},\ }\bibfield  {title} {\bibinfo {title} {Algorithm for spin symmetry operation search},\ }\href {https://journals.iucr.org/paper?ib5119} {\bibfield  {journal} {\bibinfo  {journal} {Acta Crystallogr. A Found. Adv.}\ }\textbf {\bibinfo {volume} {80}},\ \bibinfo {pages} {94} (\bibinfo {year} {2024})}\BibitemShut {NoStop}%
    \bibitem [{\citenamefont {Martin}\ and\ \citenamefont {Batista}(2008)}]{Martin2008-ze}%
      \BibitemOpen
      \bibfield  {author} {\bibinfo {author} {\bibfnamefont {I.}~\bibnamefont {Martin}}\ and\ \bibinfo {author} {\bibfnamefont {C.~D.}\ \bibnamefont {Batista}},\ }\bibfield  {title} {\bibinfo {title} {Itinerant electron-driven chiral magnetic ordering and spontaneous quantum hall effect in triangular lattice models},\ }\href {https://journals.aps.org/prl/abstract/10.1103/PhysRevLett.101.156402} {\bibfield  {journal} {\bibinfo  {journal} {Phys. Rev. Lett.}\ }\textbf {\bibinfo {volume} {101}},\ \bibinfo {pages} {156402} (\bibinfo {year} {2008})}\BibitemShut {NoStop}%
    \bibitem [{\citenamefont {Feng}\ \emph {et~al.}(2020)\citenamefont {Feng}, \citenamefont {Hanke}, \citenamefont {Zhou}, \citenamefont {Guo}, \citenamefont {Blügel}, \citenamefont {Mokrousov},\ and\ \citenamefont {Yao}}]{Feng2020-ag}%
      \BibitemOpen
      \bibfield  {author} {\bibinfo {author} {\bibfnamefont {W.}~\bibnamefont {Feng}}, \bibinfo {author} {\bibfnamefont {J.-P.}\ \bibnamefont {Hanke}}, \bibinfo {author} {\bibfnamefont {X.}~\bibnamefont {Zhou}}, \bibinfo {author} {\bibfnamefont {G.-Y.}\ \bibnamefont {Guo}}, \bibinfo {author} {\bibfnamefont {S.}~\bibnamefont {Blügel}}, \bibinfo {author} {\bibfnamefont {Y.}~\bibnamefont {Mokrousov}},\ and\ \bibinfo {author} {\bibfnamefont {Y.}~\bibnamefont {Yao}},\ }\bibfield  {title} {\bibinfo {title} {Topological magneto-optical effects and their quantization in noncoplanar antiferromagnets},\ }\href {https://www.nature.com/articles/s41467-019-13968-8} {\bibfield  {journal} {\bibinfo  {journal} {Nat. Commun.}\ }\textbf {\bibinfo {volume} {11}},\ \bibinfo {pages} {118} (\bibinfo {year} {2020})}\BibitemShut {NoStop}%
    \bibitem [{\citenamefont {Takagi}\ \emph {et~al.}(2023)\citenamefont {Takagi}, \citenamefont {Takagi}, \citenamefont {Minami}, \citenamefont {Nomoto}, \citenamefont {Ohishi}, \citenamefont {Suzuki}, \citenamefont {Yanagi}, \citenamefont {Hirayama}, \citenamefont {Khanh}, \citenamefont {Karube}, \citenamefont {Saito}, \citenamefont {Hashizume}, \citenamefont {Kiyanagi}, \citenamefont {Tokura}, \citenamefont {Arita}, \citenamefont {Nakajima},\ and\ \citenamefont {Seki}}]{Takagi2023-kk}%
      \BibitemOpen
      \bibfield  {author} {\bibinfo {author} {\bibfnamefont {H.}~\bibnamefont {Takagi}}, \bibinfo {author} {\bibfnamefont {R.}~\bibnamefont {Takagi}}, \bibinfo {author} {\bibfnamefont {S.}~\bibnamefont {Minami}}, \bibinfo {author} {\bibfnamefont {T.}~\bibnamefont {Nomoto}}, \bibinfo {author} {\bibfnamefont {K.}~\bibnamefont {Ohishi}}, \bibinfo {author} {\bibfnamefont {M.-T.}\ \bibnamefont {Suzuki}}, \bibinfo {author} {\bibfnamefont {Y.}~\bibnamefont {Yanagi}}, \bibinfo {author} {\bibfnamefont {M.}~\bibnamefont {Hirayama}}, \bibinfo {author} {\bibfnamefont {N.~D.}\ \bibnamefont {Khanh}}, \bibinfo {author} {\bibfnamefont {K.}~\bibnamefont {Karube}}, \bibinfo {author} {\bibfnamefont {H.}~\bibnamefont {Saito}}, \bibinfo {author} {\bibfnamefont {D.}~\bibnamefont {Hashizume}}, \bibinfo {author} {\bibfnamefont {R.}~\bibnamefont {Kiyanagi}}, \bibinfo {author} {\bibfnamefont {Y.}~\bibnamefont {Tokura}}, \bibinfo {author} {\bibfnamefont {R.}~\bibnamefont {Arita}}, \bibinfo {author} {\bibfnamefont {T.}~\bibnamefont
      {Nakajima}},\ and\ \bibinfo {author} {\bibfnamefont {S.}~\bibnamefont {Seki}},\ }\bibfield  {title} {\bibinfo {title} {Spontaneous topological {Hall} effect induced by non-coplanar antiferromagnetic order in intercalated van der {Waals} materials},\ }\href {https://www.nature.com/articles/s41567-023-02017-3} {\bibfield  {journal} {\bibinfo  {journal} {Nat. Phys.}\ }\textbf {\bibinfo {volume} {19}},\ \bibinfo {pages} {961} (\bibinfo {year} {2023})}\BibitemShut {NoStop}%
    \bibitem [{\citenamefont {Zhu}\ \emph {et~al.}(2024)\citenamefont {Zhu}, \citenamefont {Li}, \citenamefont {Chen}, \citenamefont {Yu},\ and\ \citenamefont {Liu}}]{Zhu2024-re}%
      \BibitemOpen
      \bibfield  {author} {\bibinfo {author} {\bibfnamefont {H.}~\bibnamefont {Zhu}}, \bibinfo {author} {\bibfnamefont {J.}~\bibnamefont {Li}}, \bibinfo {author} {\bibfnamefont {X.}~\bibnamefont {Chen}}, \bibinfo {author} {\bibfnamefont {Y.}~\bibnamefont {Yu}},\ and\ \bibinfo {author} {\bibfnamefont {Q.}~\bibnamefont {Liu}},\ }\bibfield  {title} {\bibinfo {title} {Magnetic geometry to quantum geometry nonlinear transports},\ }\href {https://www.researchgate.net/profile/Jiayu-Li-74/publication/381227013_Magnetic_geometry_to_quantum_geometry_nonlinear_transports/links/666327dda54c5f0b94565624/Magnetic-geometry-to-quantum-geometry-nonlinear-transports.pdf} {\bibfield  {journal} {\bibinfo  {journal} {arXiv [cond-mat.mtrl-sci]}\ } (\bibinfo {year} {2024})}\BibitemShut {NoStop}%
    \bibitem [{Note2()}]{Note2}%
      \BibitemOpen
      \bibinfo {note} {The spin Laue group describes the point group symmetry of SOC-free collinear magnets enhanced by the real-space inversion symmetry~\cite {Smejkal2022-zq}. After dividing the overall spin point group by the spin-only group~\cite {Litvin1974-qv}, the nontrivial spin point group is obtained as $\protect \mathcal {P} = \protect \bm {H} \cup [C_2 || A] \protect \bm {H}$ where the group $\protect \bm {H}$ consists of only the real-space operations. The spin Laue group is obtained as $\protect \bm {H} = mmm$ and $A = C_{4b}$ if one assumes the zero-$\protect \bm {Q}$ spin order in FIG.~\ref {fig:hidden_spin}\protect \,(b). The obtained group allows the $d$-wave spin splitting, implying the altermagnetic spin splitting within red- and blue-shaded cells in FIG.~\ref {fig:hidden_spin}\protect \,(b). Note that the $bc$ glide operation allowing for a mirror symmetry of $\protect \bm {H}$ is not kept due to the finite-$\protect \bm {Q}$ spin order coupled to the nonsymmorphic property.}\BibitemShut
      {Stop}%
    \bibitem [{\citenamefont {Perez-Mato}\ \emph {et~al.}(2016)\citenamefont {Perez-Mato}, \citenamefont {Gallego}, \citenamefont {Elcoro}, \citenamefont {Tasci},\ and\ \citenamefont {Aroyo}}]{Perez-Mato2016-qa}%
      \BibitemOpen
      \bibfield  {author} {\bibinfo {author} {\bibfnamefont {J.~M.}\ \bibnamefont {Perez-Mato}}, \bibinfo {author} {\bibfnamefont {S.~V.}\ \bibnamefont {Gallego}}, \bibinfo {author} {\bibfnamefont {L.}~\bibnamefont {Elcoro}}, \bibinfo {author} {\bibfnamefont {E.}~\bibnamefont {Tasci}},\ and\ \bibinfo {author} {\bibfnamefont {M.~I.}\ \bibnamefont {Aroyo}},\ }\bibfield  {title} {\bibinfo {title} {Symmetry conditions for type {II} multiferroicity in commensurate magnetic structures},\ }\href {http://dx.doi.org/10.1088/0953-8984/28/28/286001} {\bibfield  {journal} {\bibinfo  {journal} {J. Phys. Condens. Matter}\ }\textbf {\bibinfo {volume} {28}},\ \bibinfo {pages} {286001} (\bibinfo {year} {2016})}\BibitemShut {NoStop}%
    \bibitem [{\citenamefont {Corliss}\ \emph {et~al.}(1958)\citenamefont {Corliss}, \citenamefont {Elliott},\ and\ \citenamefont {Hastings}}]{Corliss1958-yi}%
      \BibitemOpen
      \bibfield  {author} {\bibinfo {author} {\bibfnamefont {L.~M.}\ \bibnamefont {Corliss}}, \bibinfo {author} {\bibfnamefont {N.}~\bibnamefont {Elliott}},\ and\ \bibinfo {author} {\bibfnamefont {J.~M.}\ \bibnamefont {Hastings}},\ }\bibfield  {title} {\bibinfo {title} {Antiferromagnetic structures of {MnS$_2$}, {MnSe$_2$}, and {MnTe$_2$}},\ }\href {https://pubs.aip.org/aip/jap/article-pdf/29/3/391/18317788/391_1_online.pdf} {\bibfield  {journal} {\bibinfo  {journal} {J. Appl. Phys.}\ }\textbf {\bibinfo {volume} {29}},\ \bibinfo {pages} {391} (\bibinfo {year} {1958})}\BibitemShut {NoStop}%
    \bibitem [{\citenamefont {Kresse}\ and\ \citenamefont {Furthmüller}(1996)}]{Kresse1996-oj}%
      \BibitemOpen
      \bibfield  {author} {\bibinfo {author} {\bibfnamefont {G.}~\bibnamefont {Kresse}}\ and\ \bibinfo {author} {\bibfnamefont {J.}~\bibnamefont {Furthmüller}},\ }\bibfield  {title} {\bibinfo {title} {Efficient iterative schemes for ab initio total-energy calculations using a plane-wave basis set},\ }\href {http://dx.doi.org/10.1103/physrevb.54.11169} {\bibfield  {journal} {\bibinfo  {journal} {Phys. Rev. B}\ }\textbf {\bibinfo {volume} {54}},\ \bibinfo {pages} {11169} (\bibinfo {year} {1996})}\BibitemShut {NoStop}%
    \bibitem [{\citenamefont {Blöchl}(1994)}]{Blochl1994-bi}%
      \BibitemOpen
      \bibfield  {author} {\bibinfo {author} {\bibfnamefont {P.~E.}\ \bibnamefont {Blöchl}},\ }\bibfield  {title} {\bibinfo {title} {Projector augmented-wave method},\ }\href {http://link.aps.org/pdf/10.1103/PhysRevB.50.17953} {\bibfield  {journal} {\bibinfo  {journal} {Phys. Rev. B}\ }\textbf {\bibinfo {volume} {50}},\ \bibinfo {pages} {17953} (\bibinfo {year} {1994})}\BibitemShut {NoStop}%
    \bibitem [{\citenamefont {Perdew}\ \emph {et~al.}(1996)\citenamefont {Perdew}, \citenamefont {Burke},\ and\ \citenamefont {Ernzerhof}}]{Perdew1996-hz}%
      \BibitemOpen
      \bibfield  {author} {\bibinfo {author} {\bibfnamefont {J.~P.}\ \bibnamefont {Perdew}}, \bibinfo {author} {\bibfnamefont {K.}~\bibnamefont {Burke}},\ and\ \bibinfo {author} {\bibfnamefont {M.}~\bibnamefont {Ernzerhof}},\ }\bibfield  {title} {\bibinfo {title} {Generalized gradient approximation made simple},\ }\href {http://dx.doi.org/10.1103/PhysRevLett.77.3865} {\bibfield  {journal} {\bibinfo  {journal} {Phys. Rev. Lett.}\ }\textbf {\bibinfo {volume} {77}},\ \bibinfo {pages} {3865} (\bibinfo {year} {1996})}\BibitemShut {NoStop}%
    \bibitem [{\citenamefont {Brostigen}\ \emph {et~al.}(1970)\citenamefont {Brostigen}, \citenamefont {Kjekshus}, \citenamefont {Liaaen-Jensen}, \citenamefont {Rasmussen},\ and\ \citenamefont {Shimizu}}]{Brostigen1970-vh}%
      \BibitemOpen
      \bibfield  {author} {\bibinfo {author} {\bibfnamefont {G.}~\bibnamefont {Brostigen}}, \bibinfo {author} {\bibfnamefont {A.}~\bibnamefont {Kjekshus}}, \bibinfo {author} {\bibfnamefont {S.}~\bibnamefont {Liaaen-Jensen}}, \bibinfo {author} {\bibfnamefont {S.~E.}\ \bibnamefont {Rasmussen}},\ and\ \bibinfo {author} {\bibfnamefont {A.}~\bibnamefont {Shimizu}},\ }\bibfield  {title} {\bibinfo {title} {Bonding schemes for compounds with the pyrite, marcasite, and arsenopyrite type structures},\ }\href {http://actachemscand.org/doi/10.3891/acta.chem.scand.24-2993} {\bibfield  {journal} {\bibinfo  {journal} {Acta Chem. Scand.}\ }\textbf {\bibinfo {volume} {24}},\ \bibinfo {pages} {2993} (\bibinfo {year} {1970})}\BibitemShut {NoStop}%
    \bibitem [{\citenamefont {Anisimov}\ \emph {et~al.}(1991)\citenamefont {Anisimov}, \citenamefont {Zaanen},\ and\ \citenamefont {Andersen}}]{Anisimov1991-na}%
      \BibitemOpen
      \bibfield  {author} {\bibinfo {author} {\bibfnamefont {V.~I.}\ \bibnamefont {Anisimov}, \bibfnamefont {VI}}, \bibinfo {author} {\bibfnamefont {J.}~\bibnamefont {Zaanen}},\ and\ \bibinfo {author} {\bibfnamefont {O.~K.}\ \bibnamefont {Andersen}},\ }\bibfield  {title} {\bibinfo {title} {Band theory and mott insulators: Hubbard \textit{U} instead of stoner {I}},\ }\href {http://link.aps.org/pdf/10.1103/PhysRevB.44.943} {\bibfield  {journal} {\bibinfo  {journal} {Phys. Rev. B}\ }\textbf {\bibinfo {volume} {44}},\ \bibinfo {pages} {943} (\bibinfo {year} {1991})}\BibitemShut {NoStop}%
    \bibitem [{\citenamefont {Persson}\ \emph {et~al.}(2006)\citenamefont {Persson}, \citenamefont {Ceder},\ and\ \citenamefont {Morgan}}]{Persson2006-tx}%
      \BibitemOpen
      \bibfield  {author} {\bibinfo {author} {\bibfnamefont {K.}~\bibnamefont {Persson}}, \bibinfo {author} {\bibfnamefont {G.}~\bibnamefont {Ceder}},\ and\ \bibinfo {author} {\bibfnamefont {D.}~\bibnamefont {Morgan}},\ }\bibfield  {title} {\bibinfo {title} {Spin transitions in the {Fe$_x$Mn$_{1-x}$S$_2$} system},\ }\href {http://link.aps.org/pdf/10.1103/PhysRevB.73.115201} {\bibfield  {journal} {\bibinfo  {journal} {Phys. Rev. B}\ }\textbf {\bibinfo {volume} {73}},\ \bibinfo {pages} {115201} (\bibinfo {year} {2006})}\BibitemShut {NoStop}%
    \bibitem [{\citenamefont {Souza}\ \emph {et~al.}(2001)\citenamefont {Souza}, \citenamefont {Marzari},\ and\ \citenamefont {Vanderbilt}}]{Souza2001-at}%
      \BibitemOpen
      \bibfield  {author} {\bibinfo {author} {\bibfnamefont {I.}~\bibnamefont {Souza}}, \bibinfo {author} {\bibfnamefont {N.}~\bibnamefont {Marzari}},\ and\ \bibinfo {author} {\bibfnamefont {D.}~\bibnamefont {Vanderbilt}},\ }\bibfield  {title} {\bibinfo {title} {Maximally localized {Wannier} functions for entangled energy bands},\ }\href {http://link.aps.org/pdf/10.1103/PhysRevB.65.035109} {\bibfield  {journal} {\bibinfo  {journal} {Phys. Rev. B}\ }\textbf {\bibinfo {volume} {65}},\ \bibinfo {pages} {035109} (\bibinfo {year} {2001})}\BibitemShut {NoStop}%
    \bibitem [{\citenamefont {Pizzi}\ \emph {et~al.}(2020)\citenamefont {Pizzi}, \citenamefont {Vitale}, \citenamefont {Arita}, \citenamefont {Blügel}, \citenamefont {Freimuth}, \citenamefont {Géranton}, \citenamefont {Gibertini}, \citenamefont {Gresch}, \citenamefont {Johnson}, \citenamefont {Koretsune}, \citenamefont {Ibañez-Azpiroz}, \citenamefont {Lee}, \citenamefont {Lihm}, \citenamefont {Marchand}, \citenamefont {Marrazzo}, \citenamefont {Mokrousov}, \citenamefont {Mustafa}, \citenamefont {Nohara}, \citenamefont {Nomura}, \citenamefont {Paulatto}, \citenamefont {Poncé}, \citenamefont {Ponweiser}, \citenamefont {Qiao}, \citenamefont {Thöle}, \citenamefont {Tsirkin}, \citenamefont {Wierzbowska}, \citenamefont {Marzari}, \citenamefont {Vanderbilt}, \citenamefont {Souza}, \citenamefont {Mostofi},\ and\ \citenamefont {Yates}}]{Pizzi2020-gd}%
      \BibitemOpen
      \bibfield  {author} {\bibinfo {author} {\bibfnamefont {G.}~\bibnamefont {Pizzi}}, \bibinfo {author} {\bibfnamefont {V.}~\bibnamefont {Vitale}}, \bibinfo {author} {\bibfnamefont {R.}~\bibnamefont {Arita}}, \bibinfo {author} {\bibfnamefont {S.}~\bibnamefont {Blügel}}, \bibinfo {author} {\bibfnamefont {F.}~\bibnamefont {Freimuth}}, \bibinfo {author} {\bibfnamefont {G.}~\bibnamefont {Géranton}}, \bibinfo {author} {\bibfnamefont {M.}~\bibnamefont {Gibertini}}, \bibinfo {author} {\bibfnamefont {D.}~\bibnamefont {Gresch}}, \bibinfo {author} {\bibfnamefont {C.}~\bibnamefont {Johnson}}, \bibinfo {author} {\bibfnamefont {T.}~\bibnamefont {Koretsune}}, \bibinfo {author} {\bibfnamefont {J.}~\bibnamefont {Ibañez-Azpiroz}}, \bibinfo {author} {\bibfnamefont {H.}~\bibnamefont {Lee}}, \bibinfo {author} {\bibfnamefont {J.-M.}\ \bibnamefont {Lihm}}, \bibinfo {author} {\bibfnamefont {D.}~\bibnamefont {Marchand}}, \bibinfo {author} {\bibfnamefont {A.}~\bibnamefont {Marrazzo}}, \bibinfo {author} {\bibfnamefont
      {Y.}~\bibnamefont {Mokrousov}}, \bibinfo {author} {\bibfnamefont {J.~I.}\ \bibnamefont {Mustafa}}, \bibinfo {author} {\bibfnamefont {Y.}~\bibnamefont {Nohara}}, \bibinfo {author} {\bibfnamefont {Y.}~\bibnamefont {Nomura}}, \bibinfo {author} {\bibfnamefont {L.}~\bibnamefont {Paulatto}}, \bibinfo {author} {\bibfnamefont {S.}~\bibnamefont {Poncé}}, \bibinfo {author} {\bibfnamefont {T.}~\bibnamefont {Ponweiser}}, \bibinfo {author} {\bibfnamefont {J.}~\bibnamefont {Qiao}}, \bibinfo {author} {\bibfnamefont {F.}~\bibnamefont {Thöle}}, \bibinfo {author} {\bibfnamefont {S.~S.}\ \bibnamefont {Tsirkin}}, \bibinfo {author} {\bibfnamefont {M.}~\bibnamefont {Wierzbowska}}, \bibinfo {author} {\bibfnamefont {N.}~\bibnamefont {Marzari}}, \bibinfo {author} {\bibfnamefont {D.}~\bibnamefont {Vanderbilt}}, \bibinfo {author} {\bibfnamefont {I.}~\bibnamefont {Souza}}, \bibinfo {author} {\bibfnamefont {A.~A.}\ \bibnamefont {Mostofi}},\ and\ \bibinfo {author} {\bibfnamefont {J.~R.}\ \bibnamefont {Yates}},\ }\bibfield  {title}
      {\bibinfo {title} {{Wannier90} as a community code: new features and applications},\ }\href {http://dx.doi.org/10.1088/1361-648X/ab51ff} {\bibfield  {journal} {\bibinfo  {journal} {J. Phys. Condens. Matter}\ }\textbf {\bibinfo {volume} {32}},\ \bibinfo {pages} {165902} (\bibinfo {year} {2020})}\BibitemShut {NoStop}%
    \bibitem [{\citenamefont {Wang}\ \emph {et~al.}(2006)\citenamefont {Wang}, \citenamefont {Yates}, \citenamefont {Souza},\ and\ \citenamefont {Vanderbilt}}]{Wang2006-sn}%
      \BibitemOpen
      \bibfield  {author} {\bibinfo {author} {\bibfnamefont {X.}~\bibnamefont {Wang}}, \bibinfo {author} {\bibfnamefont {J.~R.}\ \bibnamefont {Yates}}, \bibinfo {author} {\bibfnamefont {I.}~\bibnamefont {Souza}},\ and\ \bibinfo {author} {\bibfnamefont {D.}~\bibnamefont {Vanderbilt}},\ }\bibfield  {title} {\bibinfo {title} {\textit{Ab initio}calculation of the anomalous {Hall} conductivity by {Wannier} interpolation},\ }\href {http://link.aps.org/pdf/10.1103/PhysRevB.74.195118} {\bibfield  {journal} {\bibinfo  {journal} {Phys. Rev. B}\ }\textbf {\bibinfo {volume} {74}},\ \bibinfo {pages} {195118} (\bibinfo {year} {2006})}\BibitemShut {NoStop}%
    \bibitem [{\citenamefont {Tsirkin}\ \emph {et~al.}(2018)\citenamefont {Tsirkin}, \citenamefont {Puente},\ and\ \citenamefont {Souza}}]{Tsirkin2018-ly}%
      \BibitemOpen
      \bibfield  {author} {\bibinfo {author} {\bibfnamefont {S.~S.}\ \bibnamefont {Tsirkin}}, \bibinfo {author} {\bibfnamefont {P.~A.}\ \bibnamefont {Puente}},\ and\ \bibinfo {author} {\bibfnamefont {I.}~\bibnamefont {Souza}},\ }\bibfield  {title} {\bibinfo {title} {Gyrotropic effects in trigonal tellurium studied from first principles},\ }\href {http://link.aps.org/pdf/10.1103/PhysRevB.97.035158} {\bibfield  {journal} {\bibinfo  {journal} {Phys. Rev. B}\ }\textbf {\bibinfo {volume} {97}},\ \bibinfo {pages} {035158} (\bibinfo {year} {2018})}\BibitemShut {NoStop}%
    \bibitem [{\citenamefont {Malashevich}\ and\ \citenamefont {Souza}(2010)}]{Malashevich2010-vt}%
      \BibitemOpen
      \bibfield  {author} {\bibinfo {author} {\bibfnamefont {A.}~\bibnamefont {Malashevich}}\ and\ \bibinfo {author} {\bibfnamefont {I.}~\bibnamefont {Souza}},\ }\bibfield  {title} {\bibinfo {title} {Band theory of spatial dispersion in magnetoelectrics},\ }\href {http://link.aps.org/pdf/10.1103/PhysRevB.82.245118} {\bibfield  {journal} {\bibinfo  {journal} {Phys. Rev. B}\ }\textbf {\bibinfo {volume} {82}},\ \bibinfo {pages} {245118} (\bibinfo {year} {2010})}\BibitemShut {NoStop}%
    \bibitem [{\citenamefont {Deyo}\ \emph {et~al.}(2009)\citenamefont {Deyo}, \citenamefont {Golub}, \citenamefont {Ivchenko},\ and\ \citenamefont {Spivak}}]{Deyo2009-ah}%
      \BibitemOpen
      \bibfield  {author} {\bibinfo {author} {\bibfnamefont {E.}~\bibnamefont {Deyo}}, \bibinfo {author} {\bibfnamefont {L.~E.}\ \bibnamefont {Golub}}, \bibinfo {author} {\bibfnamefont {E.~L.}\ \bibnamefont {Ivchenko}},\ and\ \bibinfo {author} {\bibfnamefont {B.}~\bibnamefont {Spivak}},\ }\bibfield  {title} {\bibinfo {title} {Semiclassical theory of the photogalvanic effect in non-centrosymmetric systems},\ }\href {http://arxiv.org/abs/0904.1917} {\bibfield  {journal} {\bibinfo  {journal} {arXiv [cond-mat.mes-hall]}\ } (\bibinfo {year} {2009})}\BibitemShut {NoStop}%
    \bibitem [{\citenamefont {Moore}\ and\ \citenamefont {Orenstein}(2010)}]{Moore2010-sy}%
      \BibitemOpen
      \bibfield  {author} {\bibinfo {author} {\bibfnamefont {J.~E.}\ \bibnamefont {Moore}}\ and\ \bibinfo {author} {\bibfnamefont {J.}~\bibnamefont {Orenstein}},\ }\bibfield  {title} {\bibinfo {title} {Confinement-induced berry phase and helicity-dependent photocurrents},\ }\href {http://dx.doi.org/10.1103/PhysRevLett.105.026805} {\bibfield  {journal} {\bibinfo  {journal} {Phys. Rev. Lett.}\ }\textbf {\bibinfo {volume} {105}},\ \bibinfo {pages} {026805} (\bibinfo {year} {2010})}\BibitemShut {NoStop}%
    \bibitem [{\citenamefont {Sodemann}\ and\ \citenamefont {Fu}(2015)}]{Sodemann2015-vl}%
      \BibitemOpen
      \bibfield  {author} {\bibinfo {author} {\bibfnamefont {I.}~\bibnamefont {Sodemann}}\ and\ \bibinfo {author} {\bibfnamefont {L.}~\bibnamefont {Fu}},\ }\bibfield  {title} {\bibinfo {title} {Quantum nonlinear hall effect induced by berry curvature dipole in time-reversal invariant materials},\ }\href {http://dx.doi.org/10.1103/PhysRevLett.115.216806} {\bibfield  {journal} {\bibinfo  {journal} {Phys. Rev. Lett.}\ }\textbf {\bibinfo {volume} {115}},\ \bibinfo {pages} {216806} (\bibinfo {year} {2015})}\BibitemShut {NoStop}%
    \bibitem [{sup()}]{supple}%
      \BibitemOpen
      \href@noop {} {}\bibinfo {note} {See Supplemental Material}\BibitemShut {NoStop}%
    \bibitem [{\citenamefont {Newnham}(2004)}]{Newnham2004-sd}%
      \BibitemOpen
      \bibfield  {author} {\bibinfo {author} {\bibfnamefont {R.~E.}\ \bibnamefont {Newnham}},\ }\href {https://academic.oup.com/book/41813} {\emph {\bibinfo {title} {Properties of materials: Anisotropy, symmetry, structure}}}\ (\bibinfo  {publisher} {Oxford University Press},\ \bibinfo {address} {London, England},\ \bibinfo {year} {2004})\BibitemShut {NoStop}%
    \bibitem [{\citenamefont {Jaeschke-Ubiergo}\ \emph {et~al.}(2024)\citenamefont {Jaeschke-Ubiergo}, \citenamefont {Bharadwaj}, \citenamefont {Jungwirth}, \citenamefont {Šmejkal},\ and\ \citenamefont {Sinova}}]{Jaeschke-Ubiergo2024-jc}%
      \BibitemOpen
      \bibfield  {author} {\bibinfo {author} {\bibfnamefont {R.}~\bibnamefont {Jaeschke-Ubiergo}}, \bibinfo {author} {\bibfnamefont {V.~K.}\ \bibnamefont {Bharadwaj}}, \bibinfo {author} {\bibfnamefont {T.}~\bibnamefont {Jungwirth}}, \bibinfo {author} {\bibfnamefont {L.}~\bibnamefont {Šmejkal}},\ and\ \bibinfo {author} {\bibfnamefont {J.}~\bibnamefont {Sinova}},\ }\bibfield  {title} {\bibinfo {title} {Supercell altermagnets},\ }\href {http://link.aps.org/pdf/10.1103/PhysRevB.109.094425} {\bibfield  {journal} {\bibinfo  {journal} {Phys. Rev. B}\ }\textbf {\bibinfo {volume} {109}},\ \bibinfo {pages} {094425} (\bibinfo {year} {2024})}\BibitemShut {NoStop}%
    \bibitem [{\citenamefont {Ding}\ \emph {et~al.}(2020)\citenamefont {Ding}, \citenamefont {Hu}, \citenamefont {Ye}, \citenamefont {Feng}, \citenamefont {Ni},\ and\ \citenamefont {Cao}}]{Ding2020-gf}%
      \BibitemOpen
      \bibfield  {author} {\bibinfo {author} {\bibfnamefont {L.}~\bibnamefont {Ding}}, \bibinfo {author} {\bibfnamefont {C.}~\bibnamefont {Hu}}, \bibinfo {author} {\bibfnamefont {F.}~\bibnamefont {Ye}}, \bibinfo {author} {\bibfnamefont {E.}~\bibnamefont {Feng}}, \bibinfo {author} {\bibfnamefont {N.}~\bibnamefont {Ni}},\ and\ \bibinfo {author} {\bibfnamefont {H.}~\bibnamefont {Cao}},\ }\bibfield  {title} {\bibinfo {title} {Crystal and magnetic structures of magnetic topological {insulators MnBi$_2$Te$_4$ and MnBi$_4$Te$_7$}},\ }\href {https://journals.aps.org/prb/abstract/10.1103/PhysRevB.101.020412} {\bibfield  {journal} {\bibinfo  {journal} {Phys. Rev. B}\ }\textbf {\bibinfo {volume} {101}} (\bibinfo {year} {2020})}\BibitemShut {NoStop}%
    \bibitem [{\citenamefont {Blake}\ \emph {et~al.}(2005)\citenamefont {Blake}, \citenamefont {Chapon}, \citenamefont {Radaelli}, \citenamefont {Park}, \citenamefont {Hur}, \citenamefont {Cheong},\ and\ \citenamefont {Rodr\'{\i}guez-Carvajal}}]{Blake2005-of}%
      \BibitemOpen
      \bibfield  {author} {\bibinfo {author} {\bibfnamefont {G.~R.}\ \bibnamefont {Blake}}, \bibinfo {author} {\bibfnamefont {L.~C.}\ \bibnamefont {Chapon}}, \bibinfo {author} {\bibfnamefont {P.~G.}\ \bibnamefont {Radaelli}}, \bibinfo {author} {\bibfnamefont {S.}~\bibnamefont {Park}}, \bibinfo {author} {\bibfnamefont {N.}~\bibnamefont {Hur}}, \bibinfo {author} {\bibfnamefont {S.-W.}\ \bibnamefont {Cheong}},\ and\ \bibinfo {author} {\bibfnamefont {J.}~\bibnamefont {Rodr\'{\i}guez-Carvajal}},\ }\bibfield  {title} {\bibinfo {title} {{Spin structure and magnetic frustration in multiferroic RMn$_2$O$_5$ (\textit{R}=Tb,Ho,Dy)}},\ }\href {https://journals.aps.org/prb/abstract/10.1103/PhysRevB.71.214402} {\bibfield  {journal} {\bibinfo  {journal} {Phys. Rev. B}\ }\textbf {\bibinfo {volume} {71}} (\bibinfo {year} {2005})}\BibitemShut {NoStop}%
    \bibitem [{\citenamefont {Kimura}\ \emph {et~al.}(2003)\citenamefont {Kimura}, \citenamefont {Goto}, \citenamefont {Shintani}, \citenamefont {Ishizaka}, \citenamefont {Arima},\ and\ \citenamefont {Tokura}}]{Kimura2003-ma}%
      \BibitemOpen
      \bibfield  {author} {\bibinfo {author} {\bibfnamefont {T.}~\bibnamefont {Kimura}}, \bibinfo {author} {\bibfnamefont {T.}~\bibnamefont {Goto}}, \bibinfo {author} {\bibfnamefont {H.}~\bibnamefont {Shintani}}, \bibinfo {author} {\bibfnamefont {K.}~\bibnamefont {Ishizaka}}, \bibinfo {author} {\bibfnamefont {T.}~\bibnamefont {Arima}},\ and\ \bibinfo {author} {\bibfnamefont {Y.}~\bibnamefont {Tokura}},\ }\bibfield  {title} {\bibinfo {title} {{Magnetic control of ferroelectric polarization}},\ }\href {http://dx.doi.org/10.1038/nature02018} {\bibfield  {journal} {\bibinfo  {journal} {Nature}\ }\textbf {\bibinfo {volume} {426}},\ \bibinfo {pages} {55} (\bibinfo {year} {2003})}\BibitemShut {NoStop}%
    \bibitem [{\citenamefont {Katsura}\ \emph {et~al.}(2005)\citenamefont {Katsura}, \citenamefont {Nagaosa},\ and\ \citenamefont {Balatsky}}]{Katsura2005-fj}%
      \BibitemOpen
      \bibfield  {author} {\bibinfo {author} {\bibfnamefont {H.}~\bibnamefont {Katsura}}, \bibinfo {author} {\bibfnamefont {N.}~\bibnamefont {Nagaosa}},\ and\ \bibinfo {author} {\bibfnamefont {A.~V.}\ \bibnamefont {Balatsky}},\ }\bibfield  {title} {\bibinfo {title} {{Spin current and magnetoelectric effect in noncollinear magnets}},\ }\href {http://dx.doi.org/10.1103/PhysRevLett.95.057205} {\bibfield  {journal} {\bibinfo  {journal} {Phys. Rev. Lett.}\ }\textbf {\bibinfo {volume} {95}},\ \bibinfo {pages} {057205} (\bibinfo {year} {2005})}\BibitemShut {NoStop}%
    \bibitem [{\citenamefont {Jiang}\ \emph {et~al.}(2020)\citenamefont {Jiang}, \citenamefont {Nii}, \citenamefont {Arisawa}, \citenamefont {Saitoh},\ and\ \citenamefont {Onose}}]{Jiang2020-dr}%
      \BibitemOpen
      \bibfield  {author} {\bibinfo {author} {\bibfnamefont {N.}~\bibnamefont {Jiang}}, \bibinfo {author} {\bibfnamefont {Y.}~\bibnamefont {Nii}}, \bibinfo {author} {\bibfnamefont {H.}~\bibnamefont {Arisawa}}, \bibinfo {author} {\bibfnamefont {E.}~\bibnamefont {Saitoh}},\ and\ \bibinfo {author} {\bibfnamefont {Y.}~\bibnamefont {Onose}},\ }\bibfield  {title} {\bibinfo {title} {{Electric current control of spin helicity in an itinerant helimagnet}},\ }\href {http://dx.doi.org/10.1038/s41467-020-15380-z} {\bibfield  {journal} {\bibinfo  {journal} {Nat. Commun.}\ }\textbf {\bibinfo {volume} {11}},\ \bibinfo {pages} {1601} (\bibinfo {year} {2020})}\BibitemShut {NoStop}%
    \bibitem [{\citenamefont {Gäbler}\ \emph {et~al.}(2008)\citenamefont {Gäbler}, \citenamefont {Schnelle}, \citenamefont {Senyshyn},\ and\ \citenamefont {Niewa}}]{Gabler2008-uy}%
      \BibitemOpen
      \bibfield  {author} {\bibinfo {author} {\bibfnamefont {F.}~\bibnamefont {Gäbler}}, \bibinfo {author} {\bibfnamefont {W.}~\bibnamefont {Schnelle}}, \bibinfo {author} {\bibfnamefont {A.}~\bibnamefont {Senyshyn}},\ and\ \bibinfo {author} {\bibfnamefont {R.}~\bibnamefont {Niewa}},\ }\bibfield  {title} {\bibinfo {title} {Magnetic structure of the inverse perovskite ({Ce$_3$N}){In}},\ }\href {http://dx.doi.org/10.1016/j.solidstatesciences.2008.03.010} {\bibfield  {journal} {\bibinfo  {journal} {Solid State Sci.}\ }\textbf {\bibinfo {volume} {10}},\ \bibinfo {pages} {1910} (\bibinfo {year} {2008})}\BibitemShut {NoStop}%
    \bibitem [{\citenamefont {Jiang}\ \emph {et~al.}(2021)\citenamefont {Jiang}, \citenamefont {Nii}, \citenamefont {Arisawa}, \citenamefont {Saitoh}, \citenamefont {Ohe},\ and\ \citenamefont {Onose}}]{Jiang2021-yl}%
      \BibitemOpen
      \bibfield  {author} {\bibinfo {author} {\bibfnamefont {N.}~\bibnamefont {Jiang}}, \bibinfo {author} {\bibfnamefont {Y.}~\bibnamefont {Nii}}, \bibinfo {author} {\bibfnamefont {H.}~\bibnamefont {Arisawa}}, \bibinfo {author} {\bibfnamefont {E.}~\bibnamefont {Saitoh}}, \bibinfo {author} {\bibfnamefont {J.}~\bibnamefont {Ohe}},\ and\ \bibinfo {author} {\bibfnamefont {Y.}~\bibnamefont {Onose}},\ }\bibfield  {title} {\bibinfo {title} {{Chirality memory stored in magnetic domain walls in the ferromagnetic state of {MnP}}},\ }\href {http://dx.doi.org/10.1103/PhysRevLett.126.177205} {\bibfield  {journal} {\bibinfo  {journal} {Phys. Rev. Lett.}\ }\textbf {\bibinfo {volume} {126}},\ \bibinfo {pages} {177205} (\bibinfo {year} {2021})}\BibitemShut {NoStop}%
    \bibitem [{\citenamefont {Ohe}\ and\ \citenamefont {Onose}(2021)}]{Ohe2021-cb}%
      \BibitemOpen
      \bibfield  {author} {\bibinfo {author} {\bibfnamefont {J.-I.}\ \bibnamefont {Ohe}}\ and\ \bibinfo {author} {\bibfnamefont {Y.}~\bibnamefont {Onose}},\ }\bibfield  {title} {\bibinfo {title} {{Chirality control of the spin structure in monoaxial helimagnets by charge current}},\ }\href {https://doi.org/10.1063/5.0037357} {\bibfield  {journal} {\bibinfo  {journal} {Appl. Phys. Lett.}\ }\textbf {\bibinfo {volume} {118}},\ \bibinfo {pages} {042407} (\bibinfo {year} {2021})}\BibitemShut {NoStop}%
    \bibitem [{\citenamefont {Masuda}\ \emph {et~al.}(2024)\citenamefont {Masuda}, \citenamefont {Seki}, \citenamefont {Ohe}, \citenamefont {Nii}, \citenamefont {Masuda}, \citenamefont {Takanashi},\ and\ \citenamefont {Onose}}]{Masuda2024-tz}%
      \BibitemOpen
      \bibfield  {author} {\bibinfo {author} {\bibfnamefont {H.}~\bibnamefont {Masuda}}, \bibinfo {author} {\bibfnamefont {T.}~\bibnamefont {Seki}}, \bibinfo {author} {\bibfnamefont {J.-I.}\ \bibnamefont {Ohe}}, \bibinfo {author} {\bibfnamefont {Y.}~\bibnamefont {Nii}}, \bibinfo {author} {\bibfnamefont {H.}~\bibnamefont {Masuda}}, \bibinfo {author} {\bibfnamefont {K.}~\bibnamefont {Takanashi}},\ and\ \bibinfo {author} {\bibfnamefont {Y.}~\bibnamefont {Onose}},\ }\bibfield  {title} {\bibinfo {title} {{Room temperature chirality switching and detection in a helimagnetic MnAu$_2$ thin film}},\ }\href {https://www.nature.com/articles/s41467-024-46326-4} {\bibfield  {journal} {\bibinfo  {journal} {Nat. Commun.}\ }\textbf {\bibinfo {volume} {15}},\ \bibinfo {pages} {1999} (\bibinfo {year} {2024})}\BibitemShut {NoStop}%
    \bibitem [{\citenamefont {Watanabe}\ and\ \citenamefont {Yanase}(2018)}]{Watanabe2018-do}%
      \BibitemOpen
      \bibfield  {author} {\bibinfo {author} {\bibfnamefont {H.}~\bibnamefont {Watanabe}}\ and\ \bibinfo {author} {\bibfnamefont {Y.}~\bibnamefont {Yanase}},\ }\bibfield  {title} {\bibinfo {title} {{Group-theoretical classification of multipole order: Emergent responses and candidate materials}},\ }\href {https://link.aps.org/doi/10.1103/PhysRevB.98.245129} {\bibfield  {journal} {\bibinfo  {journal} {Phys. Rev. B}\ }\textbf {\bibinfo {volume} {98}},\ \bibinfo {pages} {245129} (\bibinfo {year} {2018})}\BibitemShut {NoStop}%
    \bibitem [{\citenamefont {Hayami}\ \emph {et~al.}(2018)\citenamefont {Hayami}, \citenamefont {Yatsushiro}, \citenamefont {Yanagi},\ and\ \citenamefont {Kusunose}}]{Hayami2018-bh}%
      \BibitemOpen
      \bibfield  {author} {\bibinfo {author} {\bibfnamefont {S.}~\bibnamefont {Hayami}}, \bibinfo {author} {\bibfnamefont {M.}~\bibnamefont {Yatsushiro}}, \bibinfo {author} {\bibfnamefont {Y.}~\bibnamefont {Yanagi}},\ and\ \bibinfo {author} {\bibfnamefont {H.}~\bibnamefont {Kusunose}},\ }\bibfield  {title} {\bibinfo {title} {{Classification of atomic-scale multipoles under crystallographic point groups and application to linear response tensors}},\ }\href {https://link.aps.org/doi/10.1103/PhysRevB.98.165110} {\bibfield  {journal} {\bibinfo  {journal} {Phys. Rev. B}\ }\textbf {\bibinfo {volume} {98}},\ \bibinfo {pages} {165110} (\bibinfo {year} {2018})}\BibitemShut {NoStop}%
    \bibitem [{\citenamefont {Tokura}\ \emph {et~al.}(2014)\citenamefont {Tokura}, \citenamefont {Seki},\ and\ \citenamefont {Nagaosa}}]{Tokura2014-ix}%
      \BibitemOpen
      \bibfield  {author} {\bibinfo {author} {\bibfnamefont {Y.}~\bibnamefont {Tokura}}, \bibinfo {author} {\bibfnamefont {S.}~\bibnamefont {Seki}},\ and\ \bibinfo {author} {\bibfnamefont {N.}~\bibnamefont {Nagaosa}},\ }\bibfield  {title} {\bibinfo {title} {{Multiferroics of spin origin}},\ }\href {http://dx.doi.org/10.1088/0034-4885/77/7/076501} {\bibfield  {journal} {\bibinfo  {journal} {Rep. Prog. Phys.}\ }\textbf {\bibinfo {volume} {77}},\ \bibinfo {pages} {076501} (\bibinfo {year} {2014})}\BibitemShut {NoStop}%
    \bibitem [{\citenamefont {Momma}\ and\ \citenamefont {Izumi}(2011)}]{Momma2011-jl}%
      \BibitemOpen
      \bibfield  {author} {\bibinfo {author} {\bibfnamefont {K.}~\bibnamefont {Momma}}\ and\ \bibinfo {author} {\bibfnamefont {F.}~\bibnamefont {Izumi}},\ }\bibfield  {title} {\bibinfo {title} {{VESTA} 3 for three-dimensional visualization of crystal, volumetric and morphology data},\ }\href {https://scripts.iucr.org/cgi-bin/paper?S0021889811038970} {\bibfield  {journal} {\bibinfo  {journal} {J. Appl. Crystallogr.}\ }\textbf {\bibinfo {volume} {44}},\ \bibinfo {pages} {1272} (\bibinfo {year} {2011})}\BibitemShut {NoStop}%
    \end{thebibliography}
\end{document}